\documentclass[journal=jctcce]{achemso}
\setkeys{acs}{articletitle=true}
\setkeys{acs}{etalmode=truncate}

\usepackage[english]{babel}
\usepackage[utf8]{inputenc}
\usepackage{amsmath}
\usepackage{graphicx}
\usepackage{comment}
\usepackage{dcolumn}
\usepackage{url}
\usepackage[version=3]{mhchem}

\newcolumntype{d}{D{.}{.}{4} }
\renewcommand{\arraystretch}{1.5}

\title{Minimal Basis Iterative Stockholder: Atoms in Molecules for Force-Field Development}

\author{Toon Verstraelen}
\email{toon.verstraelen@ugent.be}
\affiliation[UGent]{Center for Molecular Modeling (CMM), Member of the QCMM Ghent--Brussels Alliance, Ghent University,
Technologiepark 903, B9000 Ghent, Belgium}
\author{Steven Vandenbrande}
\affiliation[UGent]{Center for Molecular Modeling (CMM), Member of the QCMM Ghent--Brussels Alliance, Ghent University,
Technologiepark 903, B9000 Ghent, Belgium}
\author{Farnaz Heidar-Zadeh}
\affiliation[McMaster]{Department of Chemistry and Chemical Biology, McMaster University, 1280 West Main Street, Hamilton, Ontario L8S 4M1, Canada}
\author{Louis Vanduyfhuys}
\affiliation[UGent]{Center for Molecular Modeling (CMM), Member of the QCMM Ghent--Brussels Alliance, Ghent University,
Technologiepark 903, B9000 Ghent, Belgium}
\author{Veronique Van Speybroeck}
\affiliation[UGent]{Center for Molecular Modeling (CMM), Member of the QCMM Ghent--Brussels Alliance, Ghent University,
Technologiepark 903, B9000 Ghent, Belgium}
\author{Michel Waroquier}
\affiliation[UGent]{Center for Molecular Modeling (CMM), Member of the QCMM Ghent--Brussels Alliance, Ghent University,
Technologiepark 903, B9000 Ghent, Belgium}
\author{Paul W. Ayers}
\affiliation[McMaster]{Department of Chemistry and Chemical Biology, McMaster University, 1280 West Main Street, Hamilton, Ontario L8S 4M1, Canada}

\date{\today}

\begin{document}

\begin{tocentry}
    \includegraphics[viewport=0 0 250 100, width=250pt]{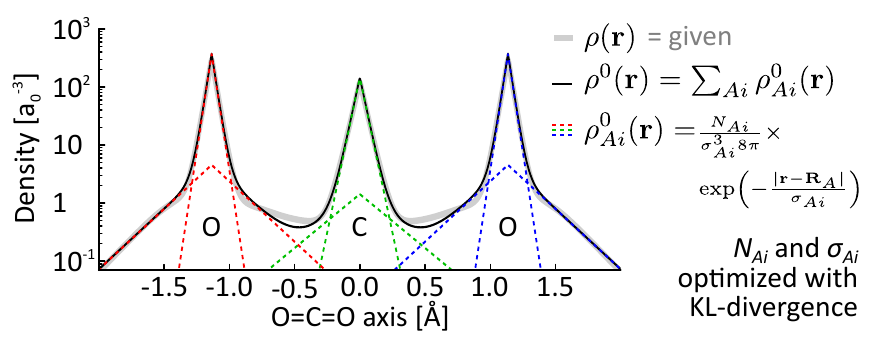}
\end{tocentry}


\begin{abstract}
Atomic partial charges appear in the Coulomb term of many force-field models and can be derived from electronic structure calculations with a myriad of atoms-in-molecules (AIM) methods. More advanced models have also been proposed, using the distributed nature of the electron cloud and atomic multipoles. In this work, an electrostatic force field is defined through a concise approximation of the electron density, for which the Coulomb interaction is trivially evaluated. This approximate ``pro-density'' is expanded in a minimal basis of atom-centered s-type Slater density functions, whose parameters are optimized by minimizing the Kullback-Leibler divergence of the pro-density from a reference electron density, e.g.\ obtained from an electronic structure calculation. The proposed method, Minimal Basis Iterative Stockholder (MBIS), is a variant of the Hirshfeld AIM method but it can also be used as a density-fitting technique. An iterative algorithm to refine the pro-density is easily implemented with a linear-scaling computational cost, enabling applications to supramolecular systems. The benefits of the MBIS method are demonstrated with systematic applications to molecular databases and extended models of condensed phases. A comparison to 14 other AIM methods shows its effectiveness when modeling electrostatic interactions. MBIS is also suitable for rescaling atomic polarizabilities in the Tkatchenko-Sheffler scheme for dispersion interactions. 
\end{abstract}

\section{Introduction}
\label{sec:intro}

The importance of force-field models is evident from recent hallmarks of atomistic force-field simulations in biology, such as the full characterization of $\beta$2 adrenergic receptor with a Markov state model \cite{kohlhoff2014} and the Anton 2 computer that can perform 10 $\mu$s molecular dynamics simulations per day on a system containing one million atoms. \cite{shaw2014} In many other domains, impressive scientific breakthroughs were also realized with atomistic force-field simulations, such as the virtual screening of $87\ 000$ zeolites for the selective adsorption of CO$_2$. \cite{kim2013}

An efficient and reliable model for electrostatic interactions is a fundamental component of a force-field model. For example, molecular recognition in proteins can be driven by electrostatic interactions. \cite{macchiarulo2003} Several authors studied partial charges derived from electronic wavefunctions to model electrostatic interactions in metal-organic frameworks. \cite{ramsahye2007, vanduyfhuys2012, kadantsev2013, gabrieli2015} Depending on the framework type, the values of the partial charges can strongly affect the predicted adsorption isotherms and self-diffusion coefficients. \cite{hamad2015} Energy decomposition methods have also shown that electrostatic interactions are one of the main driving forces in the formation of hydrogen bonds. \cite{wu2009}

In this work, we propose a new and transparent method to derive, from an electronic wavefunction, a robust, compact and reliable model for electrostatic interactions that is easily included in a force-field model. The goal is thus an efficient computation of electrostatic interactions between the molecules in the frozen density approximation, \cite{wesolowski1993, wu2009, tafipolsky2011} i.e.\ without accounting for induction or polarization effects. Although it is important and challenging to account for polarization in force fields, \cite{warshel2007, verstraelen2013_acks2, verstraelen2014} the development of polarizable force fields goes beyond the scope of this paper. We also do not consider so-called polarized force fields, where polarization is described effectively by computing the charges from an electronic structure calculation with a polarizable continuum model. \cite{ji2014}

The most basic and widespread electrostatic force-field model consists of interacting atomic point charges placed at the positions of the nuclei. In older works, e.g.\ the TraPPE force field for CO$_2$, \cite{potoff2001} the partial charges are fitted to experimental thermodynamic reference data. More recently, e.g.\ as in the TraPPE-EH models, \cite{rai2013} charges are often derived from electronic wavefunctions. Plenty of methods exist to compute such partial charges but usually, for force-field purposes, they are fitted to the electrostatic potential around model compounds of interest, \cite{fox1998} e.g.\ extensions of the AMBER force field often use the RESP method for partial charges. \cite{bayly1993} The point-charge model is only a very crude representation of the molecular charge distribution; it does not account for finer details such as atomic multipoles \cite{stone1985} and the spatial distribution of the electron cloud. The spatial distribution becomes important when electron densities of two atoms or molecules begin to overlap: in that regime point-multipole models neglect a relatively large attractive electrostatic force, which is known as the penetration effect. \cite{kairys1999, krapp2006, spackman2006, tafipolsky2011, lu2011, wang2015}

In principle, atomic multipoles are easily computed with an atoms-in-molecules (AIM) method. In some works, the acronym AIM is used exclusively for Quantum Theory of Atoms in Molecules \cite{bader1991} (QTAIM). Here it is used  more generally, to refer to any method that partitions the molecular electron density, $\rho(\mathbf{r})$, into atomic contributions, $\rho_A(\mathbf{r})$, from which e.g.\ atomic multipole moments can be derived. The spatial distribution of the electron density is sometimes also modeled with density-fitting techniques, e.g.\ as in the Gaussian Electrostatic Model \cite{piquemal2006, cisneros2006} or related methods. \cite{misquitta2014, wang2014, ohrn2016} This leads to very accurate models of the electronic density but the use of such advanced charge distributions in force-field simulations poses some difficulties: the conformational dependence of atomic multipoles can be very complex and it is far from trivial to include torques acting on higher moments in a force-field model. Several authors have proposed methods to overcome these challenges, e.g.\ with rigid molecules \cite{leslie2008} or with machine learning methods. \cite{popelier2015} Such advanced techniques are not always feasible for large-scale simulations. In this work, we propose a mathematically elegant and compact approximation of the electron distribution that results in relatively accurate electrostatic interactions in force-field models, without compromising computational efficiency. Only spherically symmetric models for atoms are considered and generalizations toward non-spherical atoms will be studied in future work.

Our new method minimizes the Kullback-Leibler (KL) Divergence of a pro-density, a minimal expansion in atom-centered s-type Slater functions, from a given molecular electron density. \cite{ghillemijn2011, heidarzadeh2015} This approach is closely related to the Iterative Stockholder (IS) method, \cite{lillestolen2008} where the pro-molecule density is a sum of spherical non-negative pro-atom densities, without any restrictions on their radial dependence. As will be discussed in section \ref{sec:mbis}, the algorithm to optimize our pro-density parameters is also very similar to IS. Hence, we refer to our new method as Minimal Basis Iterative Stockholder (MBIS).

MBIS can be perceived in two different ways. In the first place, it is a variant of the Hirshfeld method: \cite{hirshfeld1977} a partitioning of the molecular electron density inspired by information theory. \cite{nalewajski2000} Second, it can also be seen as a density fitting technique that uses the KL-divergence, \cite{heidarzadeh2015} instead of the more common least-squares approach with a Coulomb metric, \cite{elking2010} to optimize the model density. This duality permits many applications, also beyond the scope of modeling electrostatic interactions. For example, the Hirshfeld method is extensively used in different dispersion corrections for Density Functional Theory computations. \cite{becke2007, tkatchenko2009, steinmann2011} Furthermore, AIM populations are widely used in conceptual density functional theory to compute condensed reactivity indicators. \cite{geerlings2003}

Several related AIM methods were proposed in the literature, each trying to improve certain properties of their predecessors. The original Hirshfeld method \cite{hirshfeld1977} has some well-known weaknesses, such as the relatively low partial charges \cite{davidson1992} and some deficiencies in its motivation from information theory. \cite{bultinck2007}
These issues were mostly fixed in the Iterative Hirshfeld (HI) method: \cite{bultinck2007} charges computed with this method reproduce well the electrostatic potential around a molecule. \cite{vandamme2009, verstraelen2009, verstraelen2011} Compared to ESP-fitted charges, HI charges are also relatively robust with respect to conformational changes, choice of basis set, etc. \cite{bultinck2007_basis, verstraelen2011} Unfortunately, also Iterative Hirshfeld has its deficiencies. For example, when the method is applied to highly polar oxides, it requires the spherically averaged density of the non-existing oxygen dianion as input. \cite{verstraelen2013} When this dianion density is computed with a localized basis set, iterative Hirshfeld charges severely overestimate electrostatic potentials of metal oxides. \cite{verstraelen2012_silica, verstraelen2013}
The Iterative Stockholder (IS) analysis was developed independently from the Iterative Hirshfeld method and it addresses most of the issues mentioned so far. \cite{lillestolen2009, verstraelen2013} However, IS charges are not very robust with respect to conformational changes, similar to ESP-fitted charges. \cite{verstraelen2012} A recent analysis revealed that the lack of robustness is strongly related to the ill-defined density tails of the IS pro-atoms, while the core region of the IS pro-atom is usually well defined. \cite{misquitta2014} Several authors have presented solutions to overcome the weaknesses of the Iterative Hirshfeld and Iterative Stockholder methods. \cite{manz2010, manz2012, verstraelen2012, verstraelen2013, vanpoucke2013, misquitta2014} A general difficulty with these recent efforts is that they all significantly increase the algorithmic complexity and/or introduce many tuned parameters that are needed as extra input for the partitioning. In this work, we will reverse this trend and propose a method that is mathematically elegant, straightforward to implement for large systems and free from empirical input (like atomic radii) or pre-computed pro-atoms.

In the development of the MBIS method, we payed special attention to its applicability to condensed phases and extended systems. One of the applications of interest is the automatic derivation of environment specific force-field parameters for supramolecular systems \cite{cole2016} and porous materials. \cite{vanduyfhuys2012, haldoupis2012} In such applications, density partitioning is applied to DFT calculations of large atomistic models, from which force-field parameters are derived. Besides the obvious requirement that an accurate model for electrostatics must be obtained, it is also essential that the atoms-in-molecules method is computationally feasible for large systems. In practice, this means that the computational cost must scale linearly with the system size. This is achieved in MBIS by using only well-behaved integrals over atomic regions whose cost is independent of the system size.

The paper is organized as follows. In section \ref{sec:mbis}, the Minimal Basis Iterative Stockholder (MBIS) method is derived using arguments from information theory, followed by more practical aspects such as numerical algorithms and software implementations.  Section \ref{sec:example} showcases typical MBIS results with two brief applications. Section \ref{sec:comp} compares MBIS to 14 other AIM methods, assessing the robustness of charges and the accuracy of electrostatic potentials and electrostatic interactions. Some specific advantages of MBIS over (Iterative) Hirshfeld, are presented in section \ref{sec:disp}, by testing different variants of the Tkatchenko-Scheffler dispersion model. \cite{tkatchenko2009} Finally, our conclusions and an outlook on future work are given in section \ref{sec:concl}.

\section{Minimal Basis Iterative Stockholder Method}
\label{sec:mbis}

\subsection{Information theory approach to Hirshfeld partitioning}
 
It is instructive to review the information theory arguments \cite{nalewajski2000} that support the Hirshfeld method. \cite{hirshfeld1977} The amount of information lost when atoms-in-molecules (AIM) densities are approximated by pro-atoms, can be expressed as the sum of the KL-divergence for every atom:
\begin{equation}
    \label{eq:kld}
    \Delta S [\{\rho_A\};\{\rho_A^0\}]
        = \sum_{A=1}^{N_\text{atoms}} \int \rho_A(\mathbf{r}) \ln\frac{\rho_A(\mathbf{r})}{\rho_A^0(\mathbf{r})} d\mathbf{r},
\end{equation}
Traditionally, the pro-atoms, ${\rho_A^0}$, are fixed and the AIM densities, ${\rho_A}$, are the unknowns to be determined. In the original Hirshfeld method, spherically averaged isolated neutral atoms are used as pro-atoms. To obtain AIM densities that are maximally similar to the pro-atoms, one minimizes the information loss with the constraint that the AIM densities have to add up to the total density: $\sum_{A=1}^{N_\text{atoms}} \rho_A(\mathbf{r}) = \rho(\mathbf{r})$. Hence, the optimal AIM densities are a stationary point of the following Lagrangian:
\begin{equation}
\label{eq:l0}
\begin{split}
    & L_0[\{\rho_A\},\lambda(\mathbf{r});\{\rho_A^0\}] = \\
    & \qquad \sum_{A=1}^{N_\text{atoms}} \int \rho_A(\mathbf{r}) \ln\frac{\rho_A(\mathbf{r})}{\rho_A^0(\mathbf{r})} d\mathbf{r} \\
    & \quad + \int \lambda(\mathbf{r}) \left( \sum_{A=1}^{N_\text{atoms}} \rho_A(\mathbf{r}) - \rho(\mathbf{r}) \right) d\mathbf{r}.
\end{split}
\end{equation}
with the Lagrange multiplier $\lambda(\mathbf{r})$ and with fixed pro-atom densities $\rho_A^0(\mathbf{r})$. The Lagrange equations take the following form
\begin{equation}
    0 = \frac{\delta L_0}{\delta \rho_A(\mathbf{r})} = \ln \frac{\rho_A(\mathbf{r})}{\rho_A^0(\mathbf{r})} + 1 + \lambda(\mathbf{r}) \quad \forall A.
\end{equation}
The solution is:
\begin{equation}
    \frac{\rho_A(\mathbf{r})}{\rho_A^0(\mathbf{r})} = \frac{\rho_B(\mathbf{r})}{\rho_B^0(\mathbf{r})} \quad \forall A \neq B.
\end{equation}
After multiplication by $\rho_A^0(\mathbf{r}) \rho_B^0(\mathbf{r})$ and summing over all atoms $B$, one obtains the well-known stockholder partitioning: \cite{nalewajski2000}
\begin{align}
    \label{eq:stockholder}
    \rho_A(\mathbf{r}) = \rho(\mathbf{r}) \frac{\rho_A^0(\mathbf{r})}{\rho^0(\mathbf{r})}
    \quad \text{with} \quad \rho^0(\mathbf{r}) = \sum_B \rho_B^0(\mathbf{r}),
\end{align}
which corresponds to the definition originally given by Hirshfeld. \cite{hirshfeld1977} The name \textit{stockholder} comes from the ratio $\rho_A^0(\mathbf{r}) / \rho^0(\mathbf{r})$: at every point in space it represents the \textit{share} of pro-atom $A$ in the total pro-density. It can be interpreted as an atomic weight function that assigns part of the total electron density to atom $A$. In most Hirshfeld variants,\cite{hirshfeld1977, bultinck2007, lillestolen2009, verstraelen2012, verstraelen2013} the weight function varies smoothly over the range $[0,1]$. In QTAIM, \cite{bader1991} a similar atomic weight function, derived from the topology of $\rho(\mathbf{r})$, is either 1 inside the atomic basin or 0 elsewhere. Due to the minimization of the KL-divergence, the Hirshfeld AIM densities are maximally similar to the pro-atoms, ensuring some degree of transferability between AIM densities in different molecules. \cite{ayers2000}

The use of fixed pro-atoms has some important disadvantages. Results obtained with the Hirshfeld partitioning method depend largely on the choice of the fixed pro-atoms, which is essentially arbitrary. \cite{bultinck2007} Furthermore, $\rho_A^0(\mathbf{r})$ and $\rho_A(\mathbf{r})$ do not necessarily have the same norm ($N_A^0 \neq N_A$), such that the KL-divergence cannot be used as a proper measure for information loss. \cite{parr2005} This shortcoming was one of the motivations to develop the iterative Hirshfeld (HI) method. \cite{bultinck2007} In HI, the pro-atoms are not fully fixed a priori but rather updated iteratively to achieve consistency between the charge of the pro-atom and the AIM density. 

\subsection{Definition of the MBIS partitioning}

\begin{figure}
    \includegraphics{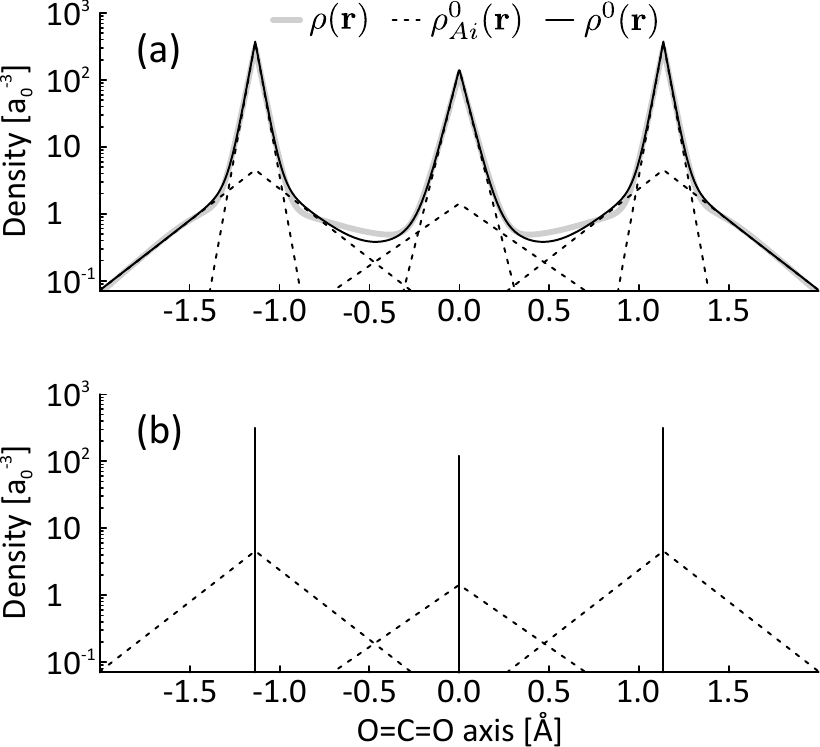}
    \caption{(a) An expansion of the molecular electron density of carbon dioxide (solid gray curve, $\rho$) in a minimal number of 1s Slater-type density functions (black dashed curves, $\rho^0_{Ai}$). The sum of all Slater functions is the pro-molecular density (solid black curve, $\rho^0$). (b) A reduction suitable for force-field models: for every atom $A$, the nuclear charge and the core Slater functions are condensed into an effective core charge (solid vertical line, $q_{A,c}$), while the valence Slater function (dashed black curve, parameters $N_{A,v}$ and $\sigma_{A,v}$) is retained.}
    \label{fig:concept}
\end{figure}

In this paper, we will make use of the information theory concepts reviewed in the previous subsection, yet with a different model for the pro-atomic density:
\begin{equation}
    \rho_A^0(\mathbf{r}) = \sum_{i=1}^{m_A} \rho_{Ai}^0(\mathbf{r}),
\end{equation}
with
\begin{equation}
    \label{eq:proslater}
    \rho_{Ai}^0(\mathbf{r}) = N_{Ai} f_{Ai}(\mathbf{r}) = \frac{N_{Ai}}{\sigma_{Ai}^3 8\pi} \exp\left(-\frac{|\mathbf{r}-\mathbf{R}_A|}{\sigma_{Ai}}\right),
\end{equation}
where the number of Slater functions, $m_{A}$, is the number of shells of atom $A$, i.e.\ its row in the periodic table. Both the population, $N_{Ai}$, and the width, $\sigma_{Ai}$, of each atomic shell are free variables. The shape functions, $f_{Ai}(\mathbf{r})$, are normalized 1s Slater-type density functions ($\int f_{Ai}(\mathbf{r}) d\mathbf{r} = 1$) and hence the population of a pro-atom is simply $N_A^0 = \sum_{i=1}^{m_A} N_{Ai}$. Figure \ref{fig:concept}a illustrates the expansion of the density in Slater functions, for the case of a carbon dioxide molecule.

It is clear that the pro-atom parameterization with s-type Slater functions is only applicable to (reconstructed) all-electron densities. Regardless of this requirement, the MBIS method has many advantages over existing methods, as will be extensively shown in the remainder of the paper. Future work will focus on more advanced pro-atom models, e.g.\ to make them also suitable for pseudo densities, while still maintaining a numerically robust algorithm. In this work, only the most minimal, yet very effective, parameterization of the pro-atoms is considered.

All the pro-atom parameters, $\{N_{Ai}\}$ and $\{\sigma_{Ai}\}$, and the AIM densities, $\rho_A(\mathbf{r})$ will be optimized by minimizing the information loss. The main difference with the conventional Hirshfeld method is that also a set of pro-atom parameters is varied, such that the pro-atom densities become a good approximation of the AIM densities. These additional degrees of freedom also allow us to constrain the population of each pro-atom and corresponding AIM to be equal, avoiding any ambiguity in the statistical interpretation of Eq.\ \eqref{eq:kld}. \cite{parr2005} The Lagrangian for this problem is an extension of Eq.\ \eqref{eq:l0} with additional variables and constraints:
\begin{equation}
\label{eq:l1}
\begin{split}
    & L_1[\{\rho_A\}, \lambda(\mathbf{r}), \{N_{Ai}\}, \{\sigma_{Ai}\}, \{\mu_A\}] = \\
    & \qquad \sum_{A=1}^{N_\text{atoms}} \int \rho_A(\mathbf{r})
             \ln\frac{\rho_A(\mathbf{r})}{\rho_A^0(\mathbf{r})} d\mathbf{r} \\
    & \quad + \int \lambda(\mathbf{r}) \left(
              \sum_{A=1}^{N_\text{atoms}} \rho_A(\mathbf{r}) - \rho(\mathbf{r})
              \right) d\mathbf{r} \\
    & \quad + \sum_{A=1}^{N_\text{atoms}} \mu_A \int \bigl(
              \rho^0_A(\mathbf{r}) - \rho_A(\mathbf{r}) \bigr) d\mathbf{r},
\end{split}
\end{equation}
where $\mu_A$ are new Lagrange multipliers associated with the consistency of the pro-atom and AIM populations.

Independent variation of the Lagrangian $L_1$ with respect to each variable ($\rho_A(\mathbf{r})$, $N_{Ai}$ or $\sigma_{Ai}$) leads to a set of Lagrange equations, which, together with the constraints, determine the MBIS AIM and pro-atom densities and the Lagrange multipliers $\lambda(\mathbf{r})$ and $\{\mu_A\}$.

We first consider the derivative of $L_1$ toward $N_{Ai}$:
\begin{align}
    0 = \frac{\partial L_1}{\partial N_{Ai}}
        & = \int \frac{\delta L_1}{\delta \rho_A^0(\mathbf{r})} \frac{\partial \rho_A^0(\mathbf{r})}{\partial N_{A,i}} d\mathbf{r} \\
        & = \int \left( -\frac{\rho_A(\mathbf{r})}{\rho_A^0(\mathbf{r})} + \mu_A \right) f_{Ai}(\mathbf{r}) d\mathbf{r} \\
        & = \mu_A - \int \frac{\rho_A(\mathbf{r})}{\rho_A^0(\mathbf{r})} f_{Ai}(\mathbf{r}) d\mathbf{r}. \label{eq:lvarn}
\end{align}
When we multiply by $N_{Ai}$ and sum over the shells $i$ of atom $A$, we get:
\begin{align}
    0 = \sum_i^{m_A} N_{Ai} \frac{\partial L_1}{\partial N_{Ai}} 
        & = \mu_A N_A^0 - \int \frac{\rho_A(\mathbf{r})}{\rho_A^0(\mathbf{r})} \sum_i^{m_A} N_{Ai} f_{Ai}(\mathbf{r}) \\
        & = \mu_A N_A^0 - N_A.
\end{align}
Due to the constraint $N_A = N_A^0$, we have $\mu_A = 1$ for each atom. Next, we take the functional derivative of $L_1$ toward $\rho_A(\mathbf{r})$ and make use of $\mu_A = 1$:
\begin{equation}
    0 = \frac{\delta L_1}{\delta \rho_A(\mathbf{r})} =
    \ln \frac{\rho_A(\mathbf{r})}{\rho_A^0(\mathbf{r})} + \lambda(\mathbf{r}),
\end{equation}
whose solution is the stockholder partitioning formula in Eq.~\eqref{eq:stockholder}. Finally, we consider the derivative of $L_1$ toward $\sigma_{Ai}$:
\begin{align}
    0 = \frac{\partial L_1}{\partial \sigma_{Ai}}
        & = \int \frac{\delta L_1}{\delta \rho_A^0(\mathbf{r})} \frac{\partial \rho_A^0(\mathbf{r})}{\partial \sigma_{A,i}} d\mathbf{r} \\
        & = -\int \frac{\rho_A(\mathbf{r})}{\rho_A^0(\mathbf{r})} \frac{\partial \rho_A^0(\mathbf{r})}{\partial \sigma_{A,i}}
         + \mu_A N_{Ai} \frac{\partial}{\partial \sigma_{Ai}} \int f_{Ai}(\mathbf{r}) d\mathbf{r} \\
        \label{eq:lvarsigma}
        & = \int \frac{\rho_A(\mathbf{r})}{\rho_A^0(\mathbf{r})} \left( \frac{3}{\sigma_{Ai}} - \frac{|\mathbf{r} - \mathbf{R}_{A}|}{\sigma_{Ai}^2} \right) \rho_{Ai}^0(\mathbf{r}) d\mathbf{r},
\end{align}
where we made use of $\int f_{Ai} (\mathbf{r}) d\mathbf{r} = 1$.

The AIM densities can be eliminated from the Lagrange equations \eqref{eq:lvarn} and \eqref{eq:lvarsigma} by making use of $\frac{\rho_A(\mathbf{r})}{\rho_A^0(\mathbf{r})} = \frac{\rho(\mathbf{r})}{\rho^0(\mathbf{r})}$. They can be rewritten in the following form:
\begin{align}
    \label{eq:updaten}
    N_{Ai} &= \int \rho(\mathbf{r})
                   \frac{\rho_{Ai}^0(\mathbf{r})}{\rho_0(\mathbf{r})}
                   d\mathbf{r}, \\
    \label{eq:updatesigma}
    \sigma_{Ai} &= \frac{1}{3N_{Ai}}
                   \int \rho(\mathbf{r})
                   \frac{\rho_{Ai}^0(\mathbf{r})}{\rho_0(\mathbf{r})}
                   |\mathbf{r} - \mathbf{R}_A|
                   d\mathbf{r}.
\end{align}
These identities form the basis for the self-consistent algorithm that will be explained in the next subsection.

\subsection{Self-consistent algorithm}

\begin{figure}
    \includegraphics{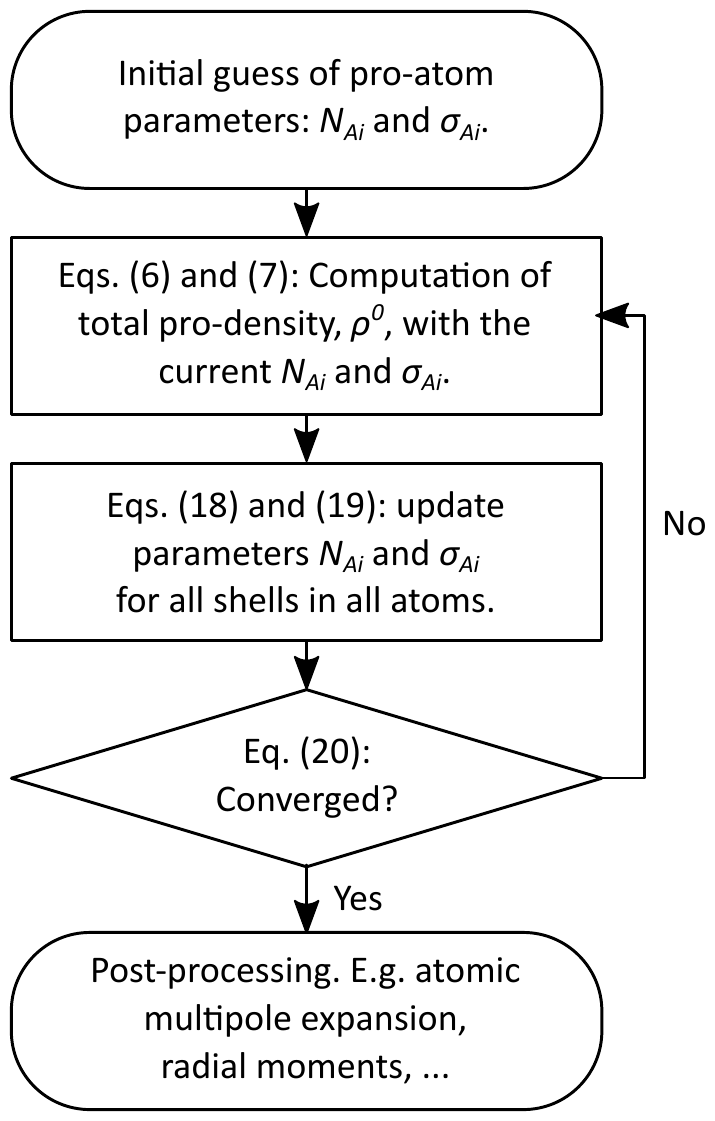}
    \caption{Flow chart of the self-consistent algorithm for the refinement of the MBIS parameters.}
    \label{fig:flowchart}
\end{figure}

Fig.\ \ref{fig:flowchart} depicts a flow chart of the self-consistent algorithm discussed in this subsection. The individual steps are described in more detail below.

In order to find all the pro-atom parameters, $\{N_{Ai}\}$ and $\{\sigma_{Ai}\}$, an initial guess is generated first, which will be refined later. Because the parameters $\{\sigma_{Ai}\}$ are non-linear, it is not guaranteed that $L_1$ is convex or has a unique minimum. Hence, multiple stationary points may exist and a reasonable initial guess is needed to find the solution of interest. The initial values of the parameters $\{N_{Ai}\}$ of atom $A$ are set to the number of electrons in each shell of the corresponding neutral isolated atom. The initial guess of $\{\sigma_{Ai}\}$ is inspired by hydrogenic s-type orbitals. For the innermost and outermost shell of atom $A$, we take $a_0/2Z_A$ and $a_0/2$, respectively, where $Z_A$ is the atomic number. Initial values of $\sigma_{Ai}$ for the intermediate shells are fixed by geometric interpolation: $\sigma_{Ai} = a_0/2 Z_A^{\left(1-\frac{i-1}{m_A-1}\right)}$.

Given the initial guess, the parameters are refined iteratively with a self-consistent update. In a single iteration, Eqs.\ \eqref{eq:updaten} and \eqref{eq:updatesigma} are evaluated, using the ``old'' parameters in the right-hand side, yielding the ``new'' parameters in the left-hand side. These iterations are repeated until the pro-atom parameters no longer change significantly. In this work, the iterative algorithm (of MBIS and other Iterative Hirshfeld flavors) is stopped after the root-mean-square deviation between the pro-atom densities of the last and the previous iteration drops below a threshold of $10^{-8} \text{ a.u}$:
\begin{equation}
    \max_A \sqrt{\int d\mathbf{r} [ \rho_{A,\text{last}}^0(\mathbf{r}) - \rho_{A,\text{previous}}^0(\mathbf{r}) ]^2} < 10^{-8}.
\end{equation}
Other convergence criteria could be used as well, e.g. based on the gradient of the Lagrangian $L_1$.

It should also be possible to optimize the pro-atom parameters with a quasi-Newton optimizer. However, robust quasi-Newton optimizers that can handle various types of equality and inequality constraints (to fix the total population and to keep all parameters positive) are non-trivial. The algorithm sketched above satisfies all constraints at every iteration, is much easier to implement and converges smoothly, even with very tight convergence settings. The Levenberg-Marquardt algorithm is not applicable because it is specifically designed for least-squares objective functions, while MBIS uses the KL-divergence as the objective function.

The integrals in Eqs.\ \eqref{eq:updaten} and \eqref{eq:updatesigma} must be evaluated numerically. Because the integrands are well localized on one atom, it is possible to implement the self-consistent update with a cost that scales linearly with the number of atoms.

\subsection{Relevant pro-atom parameters for modeling electrostatic interactions with force fields}
\label{subsec:ff}

After the optimization of pro-atom parameters, one may reduce the pro-density to a simpler picture, which is suitable for force-field models. The nuclear charge and the Slater functions associated with core electrons can be condensed into a single effective \textit{core charge}, $q_{A,c}$. This is illustrated in Fig.\ \ref{fig:concept}b. The remaining \textit{valence} Slater function is characterized by two parameters, its \textit{valence population}, $N_{A,v}$, and its \textit{valence width}, $\sigma_{A,v}$. The net atomic charge, $q_A = q_{A,c} - N_{A,v}$ can be used to approximate long-range electrostatic interactions. The two remaining degrees of freedom can be used to model the penetration effect, \cite{kairys1999, krapp2006, spackman2006, tafipolsky2011, lu2011, wang2015} i.e.\ the deviation of the electrostatic interactions from the simple point-charge model when the electronic densities begin to overlap. It can be computed efficiently with analytic expressions for the Coulomb interaction between Slater densities. \cite{lu2011, ohrn2016}

\subsection{MBIS Implementation}
\label{subsec:impl}

In the remainder of this work, MBIS will be tested extensively with applications to theoretical electron densities of molecules and condensed phases. In these applications, the all-electron density is first computed on an integration grid suitable for the numerical evaluation of Eqs.\ \eqref{eq:updaten} and \eqref{eq:updatesigma}. The implementation of these numerical integrals differs significantly between isolated molecules and periodic systems.

For isolated molecules, all-electron densities are computed with Gaussian09 \cite{g09}, using Density Functional Theory (DFT). Different functionals and Gaussian basis sets were used, as will be explained in the following sections. The MBIS partitioning of isolated molecule densities is carried out with HORTON 2.0.0, \cite{horton} which uses a standard atom-centered Becke-Lebedev integration grid. \cite{becke1988} This implementation can be combined with any level of theory in Gaussian09 that produces an all-electron 1-particle reduced density matrix (1RDM) with the ``density=current'' option. Because Gaussian09 does not write out the 1RDM when relativistic corrections are used, our tests on isolated molecules are limited to molecules containing no elements heavier than krypton.

Electron densities of periodic crystals are computed with the Projector Augmented Wave (PAW) method \cite{blochl2003} as implemented in GPAW-0.11.0. \cite{mortensen2005, enkovaara2010, bahn2002} Integrals involving the all-electron density of periodic systems are carried out as follows. In the PAW formalism, the total electron density is separated in a smoothly varying part, denoted as $\tilde{\rho}(\mathbf{r})$, and a correction for every atom $A$ in the so-called augmentation sphere. \cite{blochl2003}

In GPAW, the smooth density is represented on an equidistant real-space grid and the corrections are evaluated on atom-centered grids in spherical coordinates. This combination of integration grids makes it possible to perform very accurate integrations involving (reconstructed) all-electron densities of periodic systems. Our second MBIS implementation can handle any type of integration grid and we therefore used the same grid structure as in GPAW for periodic calculations. The advanced numerical techniques in this implementation, such as linear-scaling computational cost \cite{lee2013} and convergence acceleration, will be discussed in future work.

\subsection{Relation to other partitioning methods}

The MBIS pro-atom model has been used previously, however not yet in the context of Hirshfeld partitioning. For example, a similar pro-atom model (with fixed parameters) was also used in an ESP fitting scheme. \cite{hu2007} A similar density model is also used in the Stewart-Slater method. \cite{gill1996} Although our pro-atom model is obviously inspired by Slater's work on atomic shielding constants, \cite{slater1930} the typical polynomial prefactors are omitted. This omission is inspired by the piece-wise exponential ansatz from statistical models for atomic densities. \cite{wang1977, wang1982, fernandezpacios1991} The reduced model for force-field applications in subsection \ref{subsec:ff} has also been used before in the development of force-field models. \cite{donchev2005, wang2014, ohrn2016}

Our approach is comparable to the Iterative Stockholder (IS) method. \cite{lillestolen2009} In IS, spherical pro-atoms are defined by generic radial functions without further restrictions in terms of density basis functions; in practice they are represented by function values on a radial grid. A self-consistent update, in the same spirit as Eqs.~\eqref{eq:updaten} and \eqref{eq:updatesigma}, guarantees that the optimal IS atoms minimize the KL-divergence over all possible spherically symmetric pro-atoms. \cite{bultinck2009, lillestolen2009} Even though this is a convex problem, a well-documented weakness of IS is that the density tails of the pro-atoms are ill-defined, which leads to numerical instabilities and poorly defined atomic charges. \cite{bultinck2009, verstraelen2012, misquitta2014} In MBIS, this is resolved by modeling the density tail of each atom with only a single Slater function, which is comparable to the BS-ISA+DF method. \cite{misquitta2014} The MBIS self-consistent update algorithm is also very similar to the iterative Hirshfeld (HI) algorithm. \cite{bultinck2007} The main difference with HI is that MBIS makes use of an analytic ansatz for each pro-atom with several parameters per atom, i.e.\ the populations and widths of all shells in each atom, while HI varies just one population parameter per atom and makes use of pre-computed isolated atom densities. Furthermore, HI cannot be derived by replacing in Lagrangian $L_1$ the MBIS pro-atom by its HI counterpart. \cite{ghillemijn2011}

It is also important to realize that density fitting \cite{baerends1973, dunlap1979, fonsecaguerra1998} is closely related to MBIS. This connection becomes clear by considering the following Lagrangian:
\begin{equation}
    \label{eq:l2}
    L_2[\{N_{Ai}\}, \{\sigma_{Ai}\}, \mu]
    = \int \rho(\mathbf{r}) \ln\frac{\rho(\mathbf{r})}{\rho^0(\mathbf{r})} d\mathbf{r}
    + \mu \int \rho^0(\mathbf{r}) - \rho(\mathbf{r}) d\mathbf{r}.
\end{equation}
The self-consistent update equations \eqref{eq:updaten} and \eqref{eq:updatesigma} can also be derived from $L_2$. This shows that the optimal MBIS pro-atom parameters can also be found by minimizing the KL-divergence of the pro-molecule density, $\rho^0(\mathbf{r})$, from a given molecular density, $\rho(\mathbf{r})$. This interpretation is similar to density fitting in force-field development, \cite{piquemal2006, cisneros2006} except for the following two points. First, the MBIS pro-density is expanded in Slater functions while density-fitting techniques usually rely on contracted Gaussian functions, also with higher multipoles. \cite{elking2010} Second, MBIS uses the KL-divergence as a cost function to fit the pro-atom parameters, while conventional density-fitting makes use of a least-squares cost, often with a Coulomb metric. The least-squares cost function was also used in other related works, e.g.\ the least-squares analog of IS is known as Stewart atoms \cite{stewart1977} and the least-squares analog of MBIS is very similar to Stewart-Slater atoms. \cite{gill1996} Hybrid approaches, combining least-squares and KL-divergence cost functions, were also proposed, such as Hirshfeld-E, \cite{verstraelen2013} Gaussian ISA \cite{verstraelen2012} and BS-ISA+DF. \cite{misquitta2014} In the development of the MBIS method, a least-squares cost function was avoided because it was recently found to lead to non-local AIM densities. \cite{heidarzadeh2015} Finally, note that the pro-atom parameters are sufficient to construct monopolar electrostatic force fields. We therefore expect that the direct optimization of the pro-atoms with a Lagrangian similar to Eq.~\eqref{eq:l2}, i.e.\ without constructing AIM densities, can be an attractive alternative to conventional AIM methods.

\section{Example MBIS applications}
\label{sec:example}

This section provides two illustrative applications of the MBIS method. Their main purpose is to show the applicability of MBIS in very different scenarios and to provide the reader with some typical results. The first example discusses the robustness of MBIS and its compatibility with chemical intuition, when applied to rather extreme variations of the oxidation states of oxygen. The second example shows that MBIS is also sufficiently robust when studying subtle variations of the electron density of water between the gas, liquid and solid phase.

The examples below only illustrate the usefulness of the MBIS method. A more systematic assessment can be found in section \ref{sec:comp}.

\subsection{Oxygen in different oxidation states}
\label{subsec:oxygen}

In previous studies, Hirshfeld-I (HI) partitioning was criticized for its poor applicability to oxides. \cite{manz2010, verstraelen2012_silica, verstraelen2013} During the iterative convergence of the charges, HI requires reference densities for the oxygen dianion (or sometimes even trianion) in vacuum, which does not exist. \cite{manz2010} When the oxygen dianion is computed with a finite basis, one obtains a very diffuse density that is not representative for the oxygen atom in a molecule or crystal. This mismatch results in very large absolute values for the atomic charges in oxides, overestimating the polarity of oxide clusters or the electrostatic potential in solid oxides. \cite{manz2010, verstraelen2012_silica} Many modifications of HI were proposed to surmount this limitation, \cite{manz2010, manz2012, vanpoucke2013, verstraelen2013, bucko2013, bucko2014, gould2016} often using different (somewhat arbitrary) techniques for the computation of unstable anions. The MBIS method does not need (unstable) ion densities as input and one would therefore expect that it does not suffer from the same overpolarization issues as HI.

Table \ref{tab:examples} compares MBIS and HI results for the oxygen element, in a series of systems where the oxidation state of oxygen varies from -2 to +3: MgO, chabazite, MIL-53(Al), quartz, \ce{H2O}, \ce{CO2}, \ce{H2O2}, \ce{LiO2}, \ce{O3}, \ce{CO}, \ce{O2} and \ce{OF2}. Results for other elements in these systems are given as well for the sake of completeness. Also the isolated oxygen cation, atom and anion are included because these have different valence electron densities that result in different parameters for the outer shell in the MBIS pro-density. Atom types, which are used to differentiate all non-equivalent atoms, are defined in section S1 in the supporting information. All electron densities are computed at the PBE level of theory. For isolated molecules, Gaussian09  \cite{g09} was used with the 6-311+G(2df,p) basis and charges were computed with HORTON. \cite{horton} Electron densities of crystal unit cells were computed with GPAW, \cite{mortensen2005, enkovaara2010, bahn2002} using a grid spacing of $0.1 \textrm{ \AA}$ and charges were derived from the periodic densities with a second implementation of MBIS. (See subsection \ref{subsec:impl}.)

\renewcommand{\arraystretch}{1.0}
\begin{table}
    \centering
    \caption{Comparison of HI and MBIS charges for a selection of molecules and solids. (See text.) Results are grouped per element and sorted by the MBIS charge within each group. ON stands for oxidation number. For the MBIS method, also the quantities from subsection \ref{subsec:ff} are reported: core charge ($q_{A,c}$), valence charge ($q_{A,v}$) and valence width ($\sigma_{A,v}$).}
    \label{tab:examples}
    \begin{tabular}{llrddddd}
    \hline
     Molecule or solid & Atom$_\text{(type)}$ & ON & \multicolumn{1}{c}{HI} & \multicolumn{1}{c}{MBIS}    & \multicolumn{1}{c}{~}             & \multicolumn{1}{c}{~}             & \multicolumn{1}{c}{~}                    \\
    ~                  & ~             & ~    & \multicolumn{1}{c}{$q$ [e]} & \multicolumn{1}{c}{$q$ [e]} & \multicolumn{1}{c}{$q_{A,c}$ [e]} & \multicolumn{1}{c}{$q_{A,v}$ [e]} & \multicolumn{1}{c}{$\sigma_{A,v}$ [\AA]} \\ \hline
    MIL-53(Al)         & H$_\text{ph}$ & 0    & 0.132                       & 0.155                       & 1.0                               & -0.845                            & 0.198                                    \\
    \ce{H2O2}          & H             & 1    & 0.387                       & 0.414                       & 1.0                               & -0.586                            & 0.186                                    \\
    \ce{H2O}           & H             & 1    & 0.436                       & 0.443                       & 1.0                               & -0.557                            & 0.187                                    \\ \hline
    MIL-53(Al)         & H$_\text{hy}$ & 1    & 0.565                       & 0.519                       & 1.0                               & -0.481                            & 0.175                                    \\
    \ce{LiO2^{.}}      & Li            & 1    & 0.912                       & 0.825                       & 1.076                             & -0.251                            & 0.387                                    \\ \hline
    MIL-53(Al)         & C$_\text{pc}$ & 0    & -0.132                      & -0.148                      & 4.359                             & -4.507                            & 0.266                                    \\ \hline
    MIL-53(Al)         & C$_\text{ph}$ & 0    & -0.087                      & -0.110                      & 4.354                             & -4.463                            & 0.266                                    \\
    \ce{CO}            & C             & 3    & 0.144                       & 0.108                       & 4.327                             & -4.218                            & 0.270                                    \\
    MIL-53(Al)         & C$_\text{ca}$ & 4    & 0.910                       & 0.849                       & 4.340                             & -3.491                            & 0.246                                    \\
    \ce{CO2}           & C             & 4    & 0.847                       & 0.863                       & 4.340                             & -3.477                            & 0.243                                    \\ \hline
    \ce{MgO}           & O             & -2   & -2.220                      & -1.934                      & 6.243                             & -8.177                            & 0.247                                    \\
    Chabazite          & O$_\text{b}$  & -2   & -1.497                      & -1.261                      & 6.335                             & -7.596                            & 0.222                                    \\
    Chabazite          & O$_\text{r}$  & -2   & -1.480                      & -1.250                      & 6.337                             & -7.588                            & 0.222                                    \\
    MIL-53(Al)         & O$_\text{hy}$ & -2   & -1.952                      & -1.220                      & 6.329                             & -7.548                            & 0.223                                    \\
    Chabazite          & O$_\text{x}$  & -2   & -1.431                      & -1.219                      & 6.344                             & -7.563                            & 0.220                                    \\
    Quartz             & O             & -2   & -1.473                      & -1.213                      & 6.343                             & -7.555                            & 0.221                                    \\
    \ce{O^{-}}         & O             & -1   & -1.0                        & -1.0                        & 6.194                             & -7.194                            & 0.246                                    \\
    \ce{H2O}           & O             & -2   & -0.872                      & -0.885                      & 6.333                             & -7.219                            & 0.220                                    \\
    MIL-53(Al)         & O$_\text{ca}$ & -2   & -0.781                      & -0.748                      & 6.354                             & -7.103                            & 0.214                                    \\
    \ce{CO2}           & O             & -2   & -0.424                      & -0.431                      & 6.380                             & -6.811                            & 0.208                                    \\
    \ce{H2O2}          & O             & -1   & -0.387                      & -0.414                      & 6.351                             & -6.765                            & 0.212                                    \\
    \ce{LiO2^{.}}      & O             & -1/2 & -0.456                      & -0.412                      & 6.330                             & -6.743                            & 0.215                                    \\
    \ce{O3}            & O$_\text{t}$  & 0    & -0.194                      & -0.177                      & 6.364                             & -6.541                            & 0.207                                    \\
    \ce{CO}            & O             & -3   & -0.144                      & -0.108                      & 6.398                             & -6.506                            & 0.201                                    \\
    \ce{O}             & O             & 0    & 0.0                         & 0.0                         & 6.348                             & -6.348                            & 0.207                                    \\
    \ce{O2}            & O             & 0    & 0.0                         & 0.0                         & 6.371                             & -6.371                            & 0.203                                    \\
    \ce{OF2}           & O             & 2    & 0.156                       & 0.121                       & 6.367                             & -6.247                            & 0.203                                    \\
    \ce{O3}            & O$_\text{c}$  & 0    & 0.389                       & 0.354                       & 6.386                             & -6.032                            & 0.197                                    \\
    \ce{O^{+}}         & O             & 1    & 1.0                         & 1.0                         & 6.431                             & -5.431                            & 0.182                                    \\ \hline
    \ce{OF2}           & F             & -1   & -0.078                      & -0.060                      & 7.381                             & -7.441                            & 0.182                                    \\ \hline
    \ce{MgO}           & Mg            & 2    & 2.220                       & 1.934                       & 10.551                            & -8.617                            & 0.117                                    \\ \hline
    MIL-53(Al)         & Al            & 3    & 2.780                       & 2.111                       & 3.215                             & -1.104                            & 0.352                                    \\ \hline
    Quartz             & Si            & 4    & 2.946                       & 2.425                       & 4.425                             & -2.000                            & 0.314                                    \\
    Chabazite          & Si            & 4    & 2.944                       & 2.490                       & 4.395                             & -1.905                            & 0.316                                    \\ \hline
    \end{tabular}
\end{table}

The main trend in Table \ref{tab:examples} is the strong correlation between HI and MBIS charges. For systems where HI was found to be useful for force-field development, MBIS gives very comparable results. However, when oxygen has an oxidation state of -2 and has (semi-ionic) bonds to cations with a high oxidation number, MBIS charges for oxygen are less negative, making them more suitable for force-field development.

A reasonable correlation between atomic charges and oxidation numbers is found. Such correlations are not expected to be perfect because the oxidation number is based on simple counting rules that do not account for the (partial) covalent character of chemical bonds. It may be surprising that the core charge, $q_{A,c}$, is systematically larger than the integer value one would get by combining the nuclear charge and an integer number of core electrons. Because the valence Slater function does not decay toward the nucleus, it also contributes to the core region, which is compensated by a slightly more positive core charge. All the variations in the net charge are reflected in the valence charge, $q_{A,v}$. The valence width, $\sigma_{A,v}$, linearly correlates with the net charge: within each group of a given element, more negative atoms tend to have a slightly larger valence width.

\subsection{Application of MBIS to the three phases of water}

The MBIS method will first be illustrated with an application to an isolated water molecule, 38 clusters of water molecules, a model for the hexagonal phase of ice and 10 snapshots of a liquid water MD simulation. The electron densities of the isolated systems were computed with Gaussian09 \cite{g09} at the BLYP/6-311+G(2df,p) level of theory. \cite{lee1988, ditchfield1971} The 38 clusters, ranging from 2 to 10 water molecules in size, were taken from the work of Temelso \textit{et al}. \cite{temelso2011} We used the 3x3x2 model for the ice-1h phase of water from the work of Hayward and Reimers. \cite{hayward1997} This model contains 96 water molecules and is tuned for computational applications: the water molecules have realistic randomized orientations, yet the net dipole moment of the unit cell is constrained to zero. The geometry of the ice-1h model is refined with CP2K-2.6.0 \cite{cp2k, vandevondele2003, vandevondele2005, hutter2014} at the BLYP-D3 level of theory \cite{becke1988, lee1988, grimme2010, grimme2011} using the MOLOPT-DZVP-SR-GTH \cite{vandevondele2007} basis set and GTH pseudopotentials. \cite{goedecker1996, krack2005} CP2K was also used to generate periodic structures of liquid water (32 molecules per unit cell). These geometries were sampled every 10 ps from a 100 ps NVT \cite{bussi2007} molecular dynamics run at 300 K and at the experimental density, using the same level of theory. The electron densities of all periodic structures (ice and liquid water snapshots) were computed with GPAW \cite{mortensen2005, enkovaara2010, bahn2002} using the BLYP functional and a grid spacing of 0.1 \AA, as explained above. (CP2K was not used for this purpose because it cannot print out a reconstructed all-electron density on suitable integration grids.)

\begin{figure}
    \includegraphics{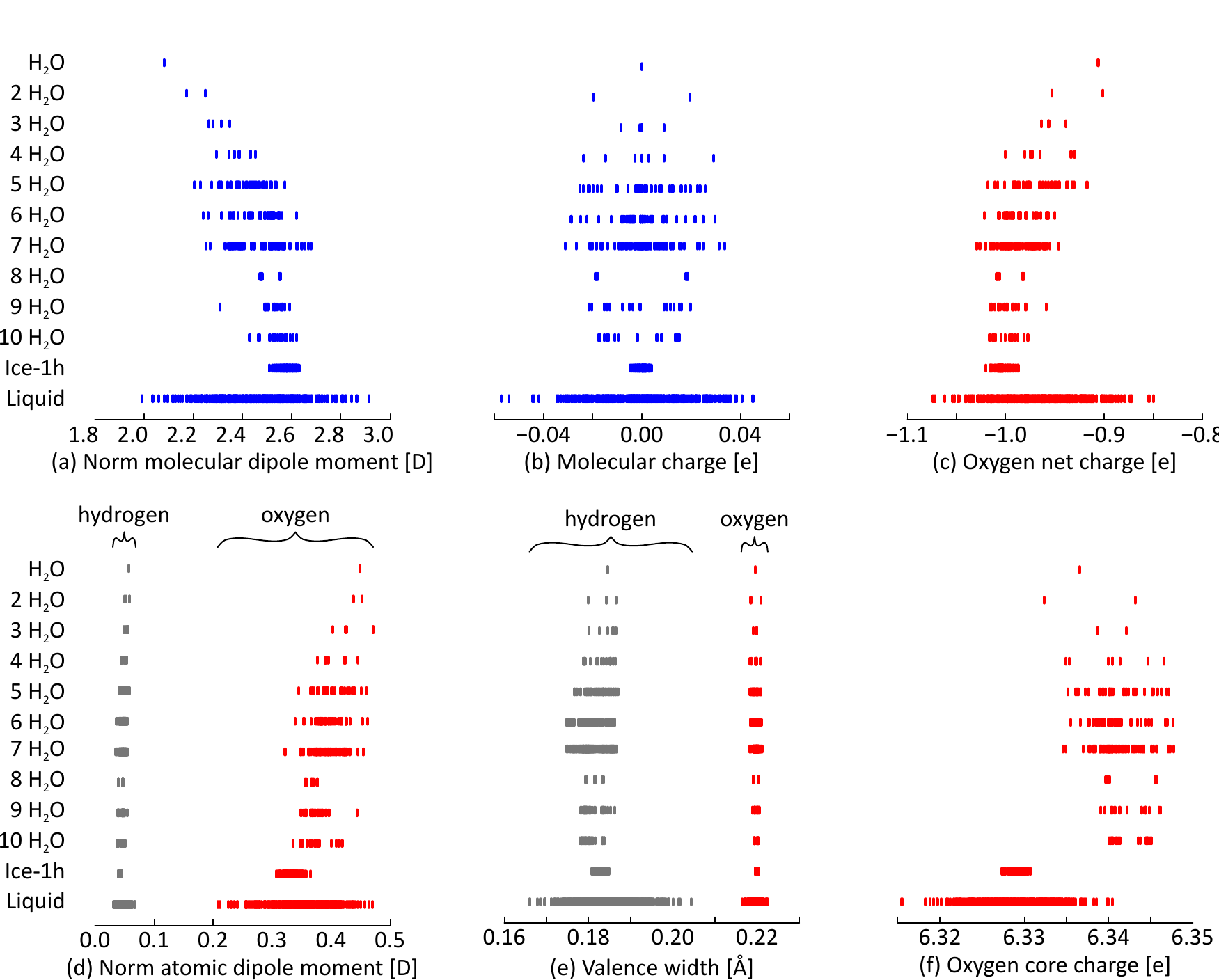}
    \caption{MBIS results for different phases of water (isolated water molecule, water clusters, ice 1h and liquid water) derived from all-electron densities computed at the BLYP level of theory: (a) the norm of the molecular dipole moment of each water molecule (b) the net charge of each water molecule (c) the oxygen charge of each water molecule, (d) the norm of each atomic dipole moment, (e) the valence width of each atom ($\sigma_{A,v}$, see subsection \ref{subsec:ff}), (f) the atomic core charge of each oxygen atom ($q_{A,c}$,  see subsection \ref{subsec:ff}).}
    \label{fig:water}
\end{figure}

Figure~\ref{fig:water} displays the key MBIS results for the water systems in this section. The partitioning of the density into atomic contributions, $\rho_A(\mathbf{r})$, is first used to construct electron densities of separate water molecules from which multipoles can be derived (relative to the molecular center of mass). The most obvious result is the increase of the molecular dipole moment as water forms hydrogen bonds with surrounding molecules (Figure~\ref{fig:water}a). Similar trends are usually found in simulations of water with polarizable force fields. \cite{horn2004, piquemal2007, burnham2008, yu2013} This increase is seen throughout all water clusters and the solid ice 1h phase. Liquid water exhibits relatively large random fluctuations in the molecular dipole moment, due to variations in the water geometry and its local environment.  Another clear trend is that water molecules tend to exchange a small fraction of an electron with their surrounding, leading to non-zero molecular charges (Figure~\ref{fig:water}b). 

Atomic charges and dipole moments are directly derived from the AIM densities, $\rho_A(\mathbf{r})$. The increased polarization of water in larger clusters (Figure~\ref{fig:water}c) is due to the decrease of the (negative) oxygen charge, while the norm of the oxygen dipole moment follows the opposite trend (Figure~\ref{fig:water}d), slightly reducing the overall polarization. Hydrogen atoms have a small and constant dipole moment, showing that they are only weakly polarizable.

Because the MBIS pro-molecular density is a sum of spherical atoms, it was to be expected that AIM dipole fluctuations play a minor role compared to atomic and molecular charge fluctuations. In general, the partitioning of the total polarization into contributions from atomic charges and/or dipoles is inherently ambiguous and can depend strongly on the AIM method. However, Mei \textit{et al.}\ observed, for a large set of molecules and for all AIM methods tested in their work, that the overall polarization always involves a significant amount of charge fluctuations. \cite{mei2015}

MBIS pro-atom parameters reveal additional trends that are not easily observed with other methods. The \textit{valence width}, $\sigma_{A,v}$ as introduced in subsection~\ref{subsec:ff}, is also sensitive to the molecular environment, which is most notable for the hydrogen atoms while oxygen has a more constant valence width (Figure~\ref{fig:water}e). Polarizable force fields usually consider fluctuating atomic charges and/or dipoles, but fluctuations in the width of the atomic electron distribution are rarely included. Our results indicate that these may also be relevant to model electronic polarization.

Finally, Figure~\ref{fig:water}f shows the \textit{atomic core charge}, $q_{A,c}$ defined in subsection~\ref{subsec:ff}. This quantity varies relatively little, in line with the expectation that the properties of core electrons should be transferable. There is a small but notable difference between the core charge for Gaussian09 (1-10 H$_2$O) and GPAW (Ice 1h, Liquid) calculations. GPAW calculations on the isolated clusters confirm that this is due to the different treatment of the core electrons in both programs (results not shown). 

In general, the MBIS results for different phases of water show that the method is robust enough to uncover several subtle trends in the electronic polarization, which is very helpful for the interpretation of these trends and the construction of polarizable force fields.  

\section{Systematic comparison to other AIM methods}
\label{sec:comp}

In this section, the MBIS method is compared to a series of other AIM methods, using several molecular datasets. This assessment focuses on properties that are relevant for modeling electrostatic interactions in force fields: the quality of the electrostatic potential (ESP), the accuracy of electrostatic interactions and the robustness of the charges.

\subsection{Molecular datasets}
\label{subsec:compmol}

Several datasets of molecular dimers and isolated molecules are considered in this section: five were taken from the literature and two new datasets are introduced below. All molecular electron densities were computed at the B3LYP/6-311+G(2df,p) level of theory with Gaussian09. \cite{g09}

Three sets of molecular dimers were taken from the work of Hobza et al., namely S66 (diverse non-covalent interactions between neutral organic molecules), \cite{rezac2011} IHB15 (ionic hydrogen bonds) \cite{rezac2012} and X40 (halogen bonds). \cite{rezac2012b} From the X40 set, dimers containing iodine were omitted because proper all-electron densities for such heavy elements can only be computed with relativistic corrections. (See subsection \ref{subsec:impl}.) Also, a new set of molecular dimers is introduced, i.e.\ ZG237, a set of 237 dimers between silica clusters and typical guest/template molecules for porous media. Neutral and anionic silica clusters are present in ZG237 and the guest molecules include noble gases, neutral and cationic organic molecules. (More details are provided in section S2 of the supporting information.) The goal of this assessment with molecular dimers is to test how well atomic charges obtained with different AIM methods can reproduce the electrostatic interaction.

\begin{figure}
    \includegraphics{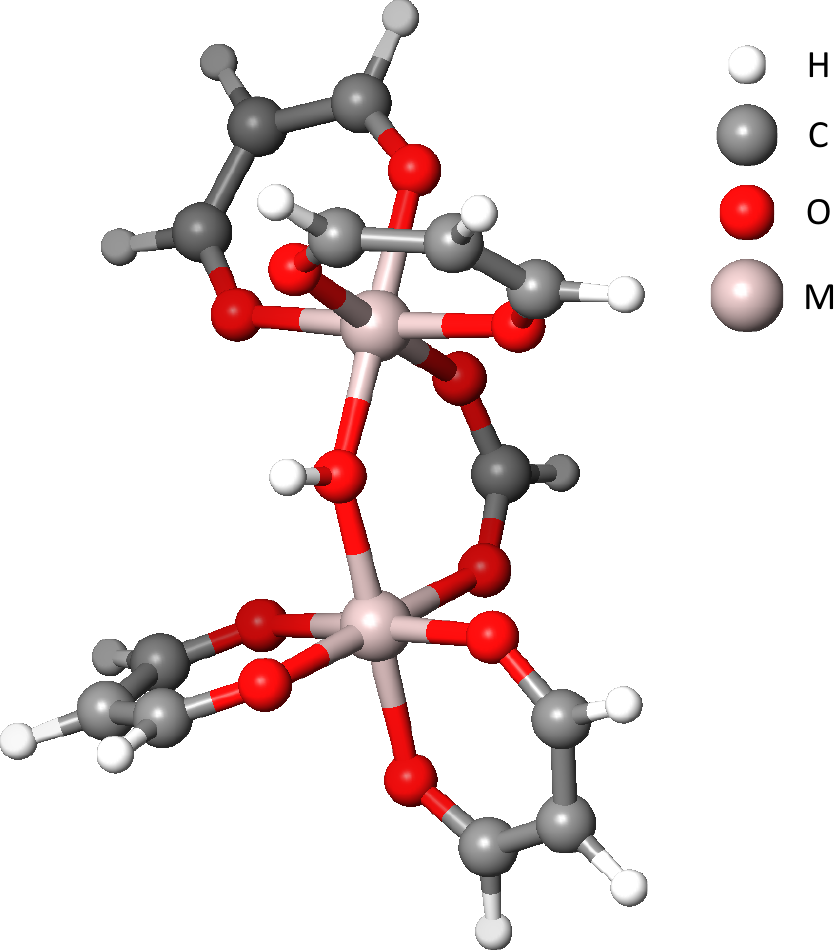}
    \caption{Metal-oxide cluster where M is a metal atom in oxidation state III (Al, Sc, Ti, V, Cr, Mn, Fe, Co, Ni, Ga).}
    \label{fig:muoh}
\end{figure}

Three datasets of larger isolated molecules are also used in the tests below, of which two were taken from earlier work: PENTA103 (103 random penta-alanine conformers) \cite{verstraelen2011} and SILICA245 (topologically different hydrogen-terminated silica clusters containing up to 8 Si atoms). \cite{verstraelen2013} One new set, MIL53(M)10 was created based on our experience with the development of a flexible force field for the metal-organic framework MIL-53(AL). \cite{vanduyfhuys2012} This set contains ten organometallic clusters with the same structure, see Fig.\ \ref{fig:muoh}, but with different metals in oxidation state III: Al, Sc, Ti, V, Cr, Mn, Fe, Co, Ni, Ga. The spin multiplicity of each cluster was fixed by coupling the spins of the transition metals to obtain a maximal $\langle \hat{S}_z \rangle$ value. The $\mu$-OH group is located at the center and the cluster is carefully terminated by four malondialdehyde anions and one formic acid anion. This neutral configuration is stable for many first-row transition metals (excluding Cu and Zn) and it resembles well the metal-oxide structure found in the MIL-53 framework. \cite{ferey2003}

\subsection{Selection of AIM methods}
\label{subsec:compaim}

Three categories of AIM methods are used for comparison: ESP-fitted charges, density partitioning methods (Hirshfeld variants and QTAIM) and Hilbert-space partitioning methods.

Atomic charges fitted to the electrostatic potential (ESP) are among the most ubiquitous for the development of force field models. We selected four such variants: Merz-Singh-Kollman (MSK), \cite{singh1984} CHELPG, \cite{breneman1990} Restrained ESP (RESP) \cite{bayly1993} and Hu-Lu-Yang (HLY). \cite{hu2007} MSK, CHELPG and HLY differ in the way the volume around the molecule is sampled in the fitting procedure but they all make use of a standard least-squares procedure. The RESP method extends the MSK cost function with hyperbolic restraints to penalize large absolute atomic charges.

A large selection of Hirshfeld variants is used in our comparison, starting with the original Hirshfeld method (H). \cite{hirshfeld1977} CM5 is a popular empirical correction to the Hirshfeld method to better reproduce experimental dipole moments of small molecules. \cite{marenich2012} Another popular related method is Hirshfeld-I (HI). \cite{bultinck2007} While the Hirshfeld method simply uses spherically averaged neutral atoms as pro-atoms, HI iteratively updates the pro-atoms to enforce consistency between the AIM charges and the charges of the pro-atoms. Charged pro-atoms are constructed by a linear interpolation between spherically averaged densities of isolated neutral atoms and ions. A particular improvement of HI over Hirshfeld is that HI charges make a good estimate of the electrostatic potential of organic molecules. \cite{vandamme2009} This is no longer the case for metal oxides, e.g.\ the ESP in the pores of zeolites, which inspired several groups to further improve the method, leading to variants such as Hirshfeld-E \cite{verstraelen2013} (HE) and DDEC4. \cite{manz2012} The Iterative Stockholder (IS) analysis is another variant of the Hirshfeld method, proposed independently of HI but with many similarities. \cite{lillestolen2008} IS also iteratively updates its pro-atoms but just uses the spherical averages of the AIM densities from the previous iteration as the new pro-atoms. Besides all the Hirshfeld variants mentioned so far (H, CM5, HI, HE, DDEC4 and IS), one more density-based method, QTAIM, \cite{bader1991} is also included in the comparison.

The third group of methods are Hilbert-space methods, which partition the density matrix instead of the density. Mulliken (M) is the oldest AIM method \cite{mulliken1955} and two popular improvements of this scheme are also widely used: L\"owdin (L) \cite{lowdin1950} and Natural (N) charges. \cite{reed1985} These three Hilbert-space methods assume that each orbital basis function is centered on one of the atoms, which is always the case when using standard Gaussian basis sets. However, in several popular periodic DFT codes, e.g.\ VASP, CPMD and GPAW, such information is not available, because they use delocalized basis sets.

Results for methods MSK, CHELPG, CM5, M, L and N were obtained with Gaussian09. \cite{g09} The MSK or CHELPG methods make use of van der Waals radii but do not define them for all elements, in which case UFF radii were used instead. Gaussian formatted checkpoint files were used to post-process the densities with HORTON-2.0.0 \cite{horton} to compute the HLY, H, HI, HE, IS and MBIS charges. The RESP program from the Antechamber program \cite{wang2006} was used to compute the RESP charges. DDEC4 charges were computed with Chargemol-09.15.2014, \cite{chargemol} QTAIM charges with AIMAll-11.06.19. \cite{aimall} Practically all MBIS results below are obtained from the pro-atom parameters discussed in section \ref{subsec:ff}. MBIS AIM densities as such are not used unless noted otherwise.

\subsection{Quality of the electrostatic potential (ESP)}
\label{subsec:compesp}

In the context of force-field development, it is assumed that whenever atomic charges accurately reproduce the electrostatic potential (ESP) around a molecule, they also make good predictions of the electrostatic interactions. \cite{fox1998} Hence, one of the desirable properties of atomic charges is their ability to reproduce the ESP as well as possible, which is the topic of this subsection. A direct assessment of the quality of electrostatic interactions is discussed in subsection \ref{subsec:compint}.

We have tested the quality of the ESP for all sets of isolated molecules discussed in section \ref{subsec:compmol} and also for the monomers present in all dimer datasets. The HLY ESP cost function is used to measure the quality of the ESP and the results would not change much if we had used an MSK or CHELPG cost function instead. The HLY cost functions is an integration over a volume surrounding the molecule, \cite{hu2007} which we converted to an RMSE value as follows:
\begin{equation}
    \label{eq:rsmdesp}
    \text{RMSE}_\text{ESP} = \sqrt{\frac{
        \int w_\text{HLY}(\mathbf{r}) \left(
            V_\text{DFT}(\mathbf{r}) 
            - \sum_A \frac{q_A}{4 \pi \epsilon_0|\mathbf{r} - \mathbf{R}_A|}
        \right)^2 d\mathbf{r}
    }{
        \int w_\text{HLY}(\mathbf{r}) d\mathbf{r}
    }}
\end{equation}
where $w_\text{HLY}(\mathbf{r})$ is the weight function designed by Hu, Lu and Yang: it becomes one in the region surrounding a molecule and goes smoothly to zero inside the molecule and at larger distances. \cite{hu2007} $V_\text{DFT}(\mathbf{r})$ is the reference ESP from the DFT calculation and $q_A$ are the atomic charges. The smoothness of the weight function guarantees that the ESP cost is not sensitive to the exact position of the grid points, which is a clear advantage over other ESP fitting methods. RMSE$_\text{ESP}$, which we computed for every molecule and every AIM method, is a measure for the error on the frozen-density interaction energy \cite{wesolowski1993, wu2009, tafipolsky2011} of a unit charge with the molecule when it is placed near its van der Waals surface.

\begin{figure}
    \includegraphics{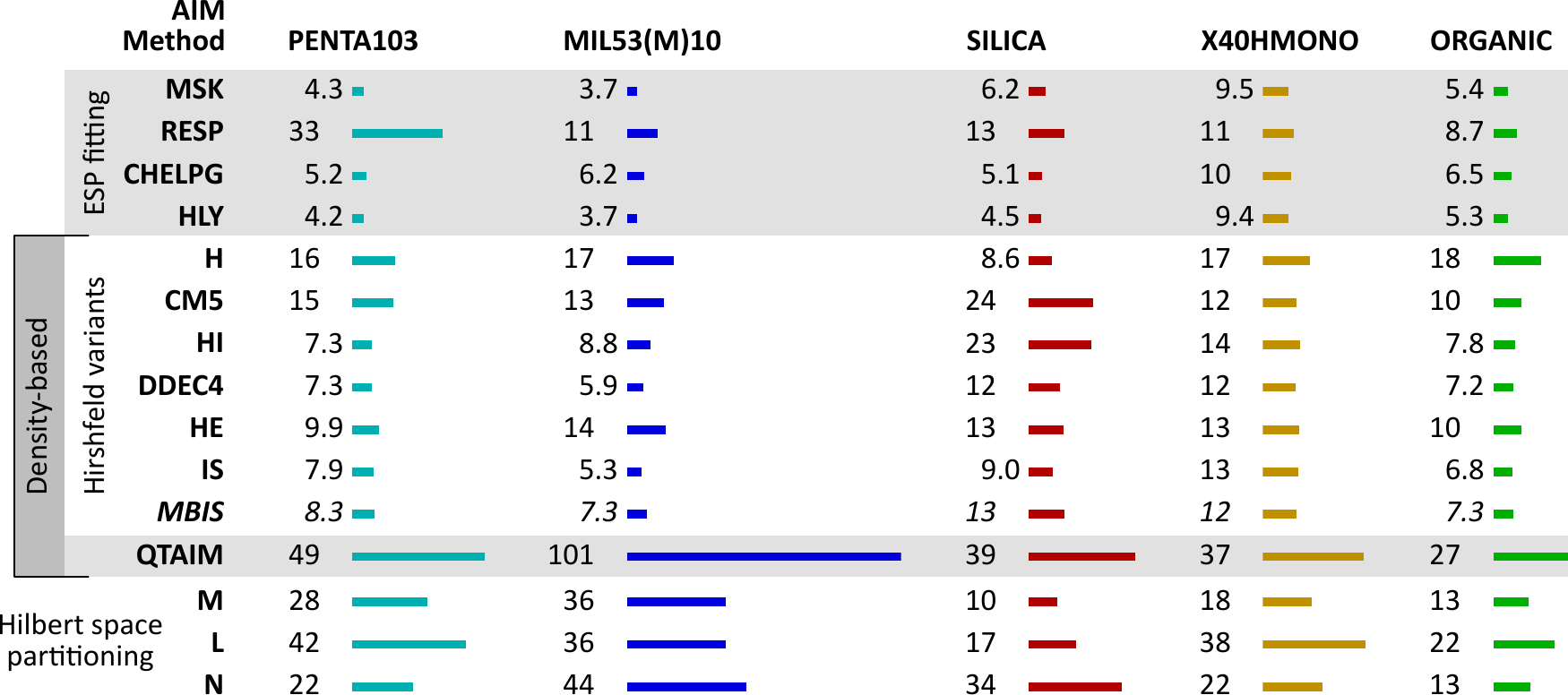}
    \caption{RMSE$_\text{ESP}$ in kJ\,mol$^{-1}$ computed for all AIM methods in this work, averaged over groups of isolated molecules. (See text for definition of groups).}
    \label{fig:esptab}
\end{figure}

Fig.\ \ref{fig:esptab} compares for every AIM method the average of RMSE$_\text{ESP}$ within five groups of isolated molecules: PENTA103 and MIL53(M)10 are those discussed in section \ref{subsec:compmol}. SILICA contains all those of the SILICA245 set plus all silica clusters from the ZG237 set. X40HMONO contains all halogenide molecules present in the X40 set of dimers. Finally, ORGANIC contains all other monomers from the dimer sets S66, IHB15, X40 and ZG237. (The noble gas atoms from ZG237 are not included.)

Obviously, the ESP-fitted charges (MSK, RESP, CHELPG and HLY) perform well in this test as they are optimized to reproduce the ESP surrounding each molecule. RESP charges are not as optimal as the other three because of the hyperbolic restraints, which becomes very pronounced for large molecules as in the PENTA set. Lowering the strength of the restraints could relieve this issue but it would also result in less robust charges. The halogenides in X40HMONO have an ESP that is relatively difficult to reproduce with point charges: the sigma-hole of the halogen atom corresponds to a large and local dipole moment whose effect on the ESP cannot be explained in terms of atomic monopoles. \cite{ibrahim2011, cole2016}

The original Hirshfeld method (H) usually predicts poor ESPs because the absolute values of the atomic charges are too low. \cite{davidson1992} All variants of the Hirshfeld method (HI, CM5, DDEC4, HE, IS and MBIS) produce more accurate ESPs, except for CM5 and HI when tested with the SILICA set. The good performance of the IS method is not surprising: it partitions the electron density in AIMs that are as spherical as possible, sometimes by introducing an unreasonable radial dependence, \cite{verstraelen2009} thus having small atomic dipole and higher multipole moments. In fact, any method that performs better than IS likely biases the charges to mimic effects of atomic multipoles. Such overfitting clearly occurs in the ESP-fitting methods. In case of MBIS, we only considered MBIS point charges and not the more advanced model with valence Slater functions, see Fig.\ \ref{fig:concept}b, simply because Eq.\ \eqref{eq:rsmdesp} only tests the ESP outside the molecule where the density is very low. In this region, the ESP generated by the Slater functions is very well approximated by that of point charges. Of all Hirshfeld variants, DDEC4, IS and MBIS are comparably good.

In line with previous observations, QTAIM charges are inadequate for the purpose of modeling ESPs. \cite{verstraelen2013} This can only be fixed by including higher atomic QTAIM multipoles, as is often done in QTAIM-based force fields. \cite{popelier2015} Mulliken (M), L\"owdin (L) and Natural (N) charges rarely produce useful ESPs in our tests.

\begin{figure}
    \includegraphics{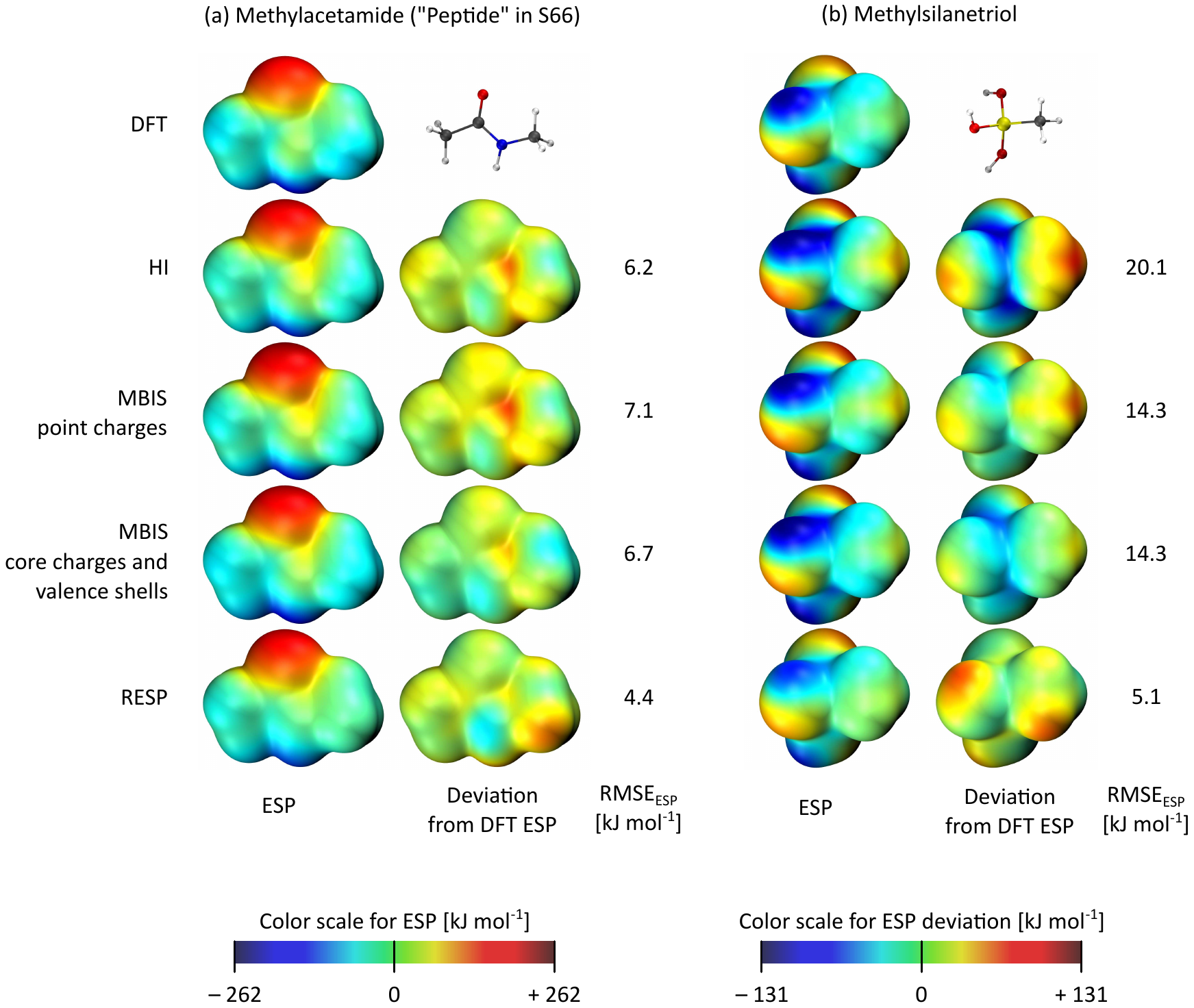}
    \caption{ESP maps plotted on the $\rho=0.002 \text{ a.u.}$ isosurface of (a) methylacetamide and (b) methylsilanetriol. In the left column for each molecule, the DFT ESP and the ESP due to selected point charge methods (HI, MBIS and RESP) and the core charge + valence shell (MBIS) are shown. For each approximate ESP, the deviation from the DFT reference is shown in the right column, for each molecule.}
    \label{fig:isosurfaces}
\end{figure}

Fig.\ \ref{fig:isosurfaces} shows ESP maps plotted on the $\rho=0.002 \text{ a.u.}$ isosurface of two representative molecules: methylacetamide and methylsilanetriol, which have an electrostatic potential that is respectively easy and difficult to reproduce with point charges. The isosurface approximates the molecular van der Waals surface, \cite{bader1987} which is convenient for visualizing non-covalent interactions. In addition to the ESP of the DFT calculation, the ESPs obtained with a subset of atomic charge methods, and their deviation from the DFT result, are shown. The isosurfaces sample the ESP at a higher density than the HLY cost function, which has two important consequences. First, the scale of the ESP deviations is large compared to the reported RMSE$_\text{ESP}$ values. Second, the ESP maps are different for MBIS point charges and MBIS core charges with delocalized valence shells, the latter accounting for the penetration effect.

The main observation is that all model ESPs qualitatively agree with the DFT result. (See left column in Fig.\ \ref{fig:isosurfaces}a and Fig.\ \ref{fig:isosurfaces}b.) Some quantitative differences are present but they only appear clearly in the isosurfaces on which the deviations from the DFT ESP are shown. (See right column in Fig.\ \ref{fig:isosurfaces}a and Fig.\ \ref{fig:isosurfaces}b.) Even though RESP charges have a relatively low RMSE$_\text{ESP}$, the deviations from the DFT ESP are not significantly smaller than for the other methods. When MBIS core charges and valence shells are used to estimate the ESP, a better visual agreement is found, because the penetration effect is already significant at the selected isodensity surface. Also note that the ESP obtained with HI charges for methylsilanetriol deviates the most from the DFT reference, in line with the limitations of HI for oxides, which were also discussed in subsection \ref{subsec:oxygen} and which is also seen in Fig.\ \ref{fig:esptab}. Finally, note that this visualization of two representative molecules merely serves as an illustration. Solid conclusions can only be drawn from a thorough statistical analysis involving many molecules, such as the one presented in Fig.\ \ref{fig:esptab}.

\subsection{Accuracy of electrostatic interactions}
\label{subsec:compint}

A common assumption in force-field development is that ESP-fitted charges also reproduce electrostatic interaction energies in general. Here, we assess the validity of this assumption for molecular dimers: the electrostatic interaction in the frozen-density approximation, \cite{wesolowski1993, wu2009, tafipolsky2011} $E_\text{FD}$, will be used as a reference to test approximate electrostatic interactions obtained with atomic point charges from different AIM methods. The frozen-density approximation does not include any effects from polarization or charge-transfer. Such effects should be modeled with a polarizable (or polarized) force field, which is beyond the scope of this test.

\begin{figure}
    \includegraphics{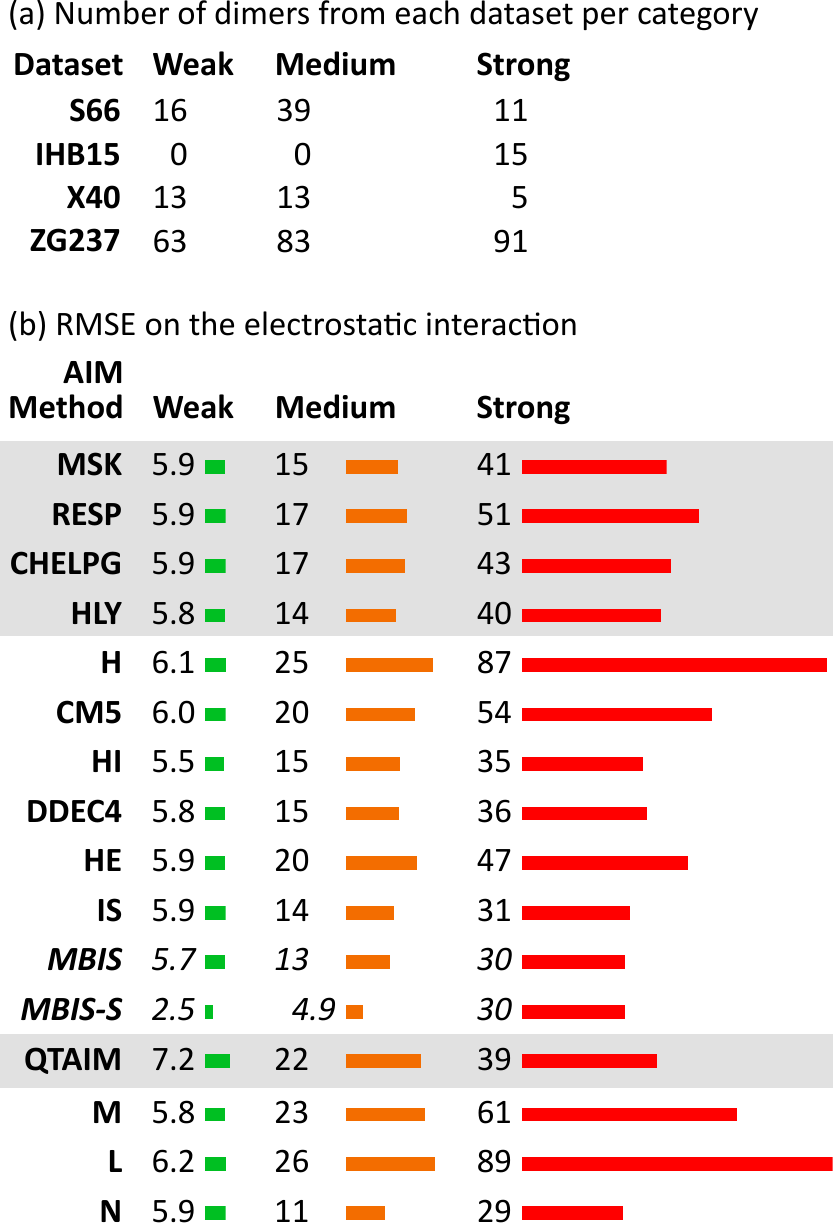}
    \caption{The accuracy of the electrostatic interactions for three classes of electrostatic interactions: Weak ($E_\text{FD} > -10\,\text{kJ\,mol}^{-1}$),  Medium electrostatics ($-10\,\text{kJ\,mol}^{-1} \ge E_\text{FD} > -50\,\text{kJ\,mol}^{-1}$) and Strong ($E_\text{FD} \le -50\,\text{kJ\,mol}^{-1}$). Part (a) shows the number of dimers each dataset contributes to each class. Part (b) shows the RMSE of the electrostatic interaction in kJ\,mol$^{-1}$, obtained with atomic charges from different AIM methods, for each group. MBIS-S goes beyond the simple point-charge model and uses spherical valence Slater functions to model the penetration effect.}
    \label{fig:inttab}
\end{figure}

The four sets of molecular dimers described in section \ref{subsec:compmol} (S66, IHB15, X40 and ZG237) cover a large variety of electrostatic interactions, from as little as $-0.06$ kJ\,mol$^{-1}$ to rather extreme values of $-664$ kJ\,mol$^{-1}$. Especially in the ZG237 set, it is often hard to classify dimers into specific interaction types, like hydrogen bonding, salt bridge, etc. To facilitate the interpretation of the results, we have classified the dimers more conveniently, just using thresholds on the strength of the electrostatic interaction in the frozen-density approximation: ``Weak'' ($E_\text{FD} > -10\,\text{kJ\,mol}^{-1}$),  ``Medium'' ($-10\,\text{kJ\,mol}^{-1} \ge E_\text{FD} > -50\,\text{kJ\,mol}^{-1}$) and ``Strong'' ($E_\text{FD} \le -50\,\text{kJ\,mol}^{-1}$). Fig.\ \ref{fig:inttab}a shows the numbers of dimers from each dataset in each class. The IHB15 set contributes exclusively to the ``Strong'' class while all other sets have dimers in each class of interaction strength. Fig.\ \ref{fig:inttab}b shows the root-mean-square error (RMSE) on the electrostatic interaction energy for each of the three classes and for each AIM method. The label ``MBIS-S`` refers to the interaction energy computed using effective core charges and valence Slater density functions, as shown schematically in Fig.\ \ref{fig:concept}b.

A surprising result in Fig.\ \ref{fig:inttab} is that electrostatic interactions computed with ESP-fitted charges are not the most accurate. MBIS point charges perform better in all three classes (Weak, Medium and Strong) than the best ESP-fitting method (HLY). This can be understood as follows: as explained in subsection \ref{subsec:compesp}, ESP-fitted charges are biased to reproduce effects of atomic multipoles on the ESP. Although this may improve the accuracy with which the ESP is reproduced, it is a form of overfitting that may deteriorate other results obtained with ESP-fitted charges, as we observe here.

The second important result is that the MBIS method is a very effective model for the penetration effect. The MBIS-S results in Fig.\ \ref{fig:inttab} are obtained by describing every atom with an effective core charge and a valence Slater function. Even though this is a very simple (and thus computationally efficient) approach, it already reduces the RMSE by more than 50\% in the classes ``Weak'' and ``Medium''. Only for ``Strong'' electrostatic interactions, there are no apparent benefits from using such Slater density functions. A more detailed analysis, in which we computed electrostatic interactions with multipole expansions of MBIS AIM densities, showed that the largest error on the ``Strong`` electrostatic interactions is due to the neglect of atomic dipole moments.

\subsection{Robustness of the atomic charges}
\label{subsec:comprob}

For the development of an electrostatic force-field model or for the chemical interpretation of atomic charges, it is desirable that the charges are robust, i.e.\ not too sensitive to small details in the electronic structure calculations from which they are derived. Robustness is a prerequisite for transferability, i.e.\ the assumption that parameters derived from a molecule remain valid when that molecule is embedded (non-)covalently in a molecular environment. Even for environment-specific force-field parameters, \cite{grimme2014, cole2016} a robust partitioning is of interest to assure that such parameters remain valid as far as possible from the reference point for which they were computed. In this subsection, three kinds of sensitivity (the inverse of robustness) of atomic charges are investigated: sensitivity to conformational changes, to chemical changes in the environment and to changes in the basis set.

\begin{figure}
    \includegraphics{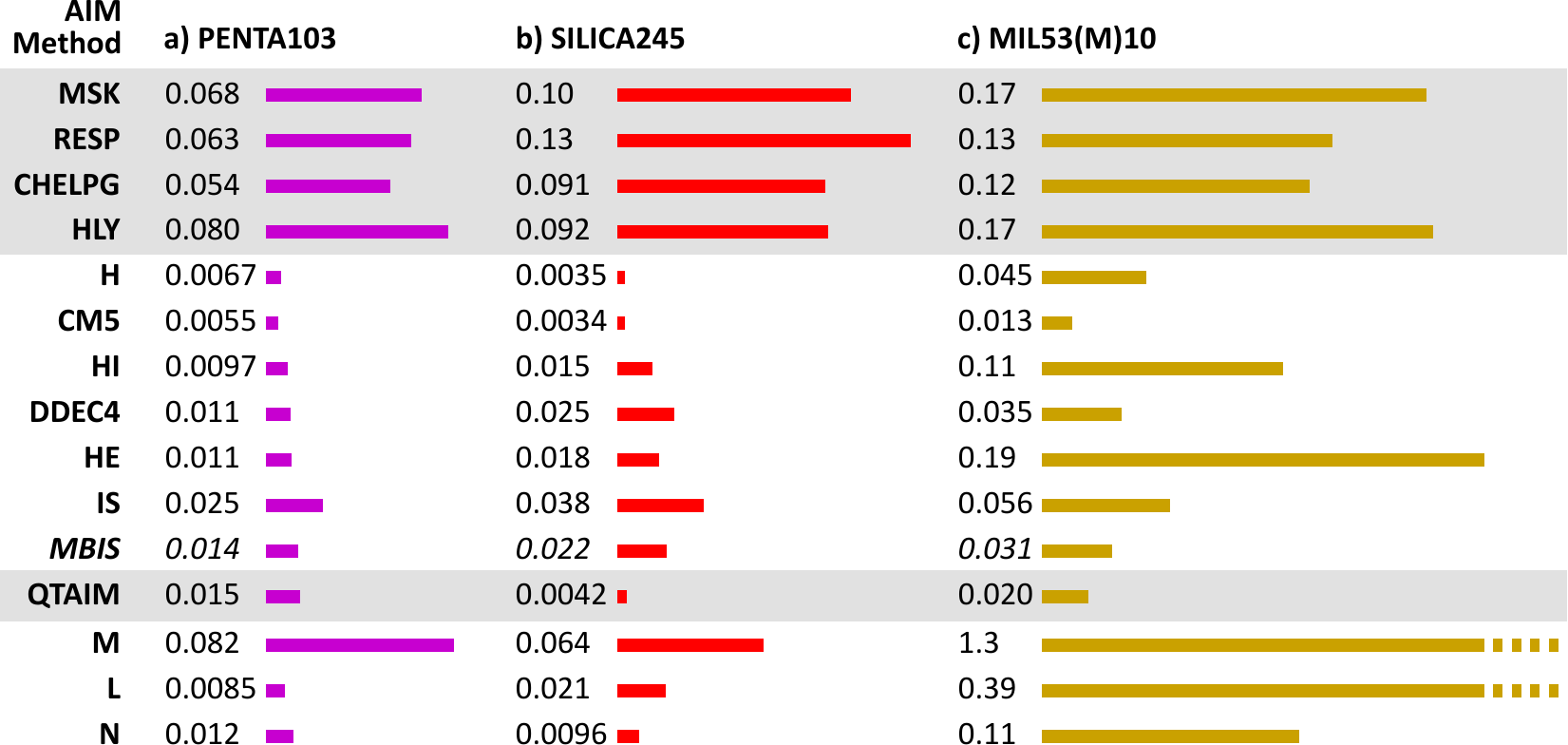}
    \caption{Standard deviations on atomic charges in e, within different sets of molecules. (a) The fluctuation on the atomic charges of 103 penta-alanine conformers. (b) The fluctuation on three different types of Si charges in a set if 245 silica clusters. (See text for definition of groups of Si atoms.) (c) The fluctuation on the metal charge in the MIL53(M)10 clusters, due to changes in basis set.}
    \label{fig:senstab}
\end{figure}

Fig.\ \ref{fig:senstab}a shows the sensitivity of the atomic charges to conformational changes of the penta-alanine chain, for all AIM methods. For a given AIM method, the standard deviation of the atomic charges in the PENTA103 set are computed with respect to the average charge of each atom over all 103 conformations. These 103 conformers are randomly generated meta-stable structures. \cite{verstraelen2011} Although some fluctuation of the charges may be expected due to internal polarization, some methodological artifacts will cause larger fluctuations without physical origin.

Fig.\ \ref{fig:senstab}b compares an averaged standard deviation of the Si charges in the SILICA245 set. The Si atoms are divided into three groups, based on the number of terminating hydrogen atoms they are bonded to (ranging from 1 to 3). Within each group, the standard deviation on the Si charge is computed and the average over the three groups is shown in Fig. \ref{fig:senstab}b. This standard deviation should be small because the Si atoms within one group have a very comparable chemical environment.

Finally, Fig.\ \ref{fig:senstab}c shows the sensitivity of the metal atom charge in the MIL53(M)10 clusters to the basis set. Their electron densities were computed with 6-311+G(2dp,f), 6-311+G*, 6-31+G*, 6-31+G or 6-31G*. The standard deviation is computed relative to the average charge of each transition metal over all basis sets. Note that the sensitivity values for Mulliken and L\"owdin fall literally off the chart and the corresponding bars in the bar plot were truncated for the sake of clarity.

The ESP-fitted and Mulliken charges have a very high sensitivity in all three cases, in line with earlier work. \cite{verstraelen2011} This is problematic because it is almost impossible to provide definitive charges with such methods. The basis set sensitivity seems to be the most difficult to control: the standard deviation is larger than 0.1 e for the methods MSK, RESP, CHELPG, HLY, HI, HE, and M, L and N.

The basis set robustness of H, HI and HE can be improved as follows. Currently we used consistent levels of theory for pro-atom and molecular electron densities. If the pro-atoms were computed with a single level of theory and basis set, independent of the settings of the molecular calculation, the robustness would significantly improve. This is noticeable in the low sensitivity of the DDEC4 and especially the CM5 charges. Both CM5 and DDEC are implemented with a unique set of pro-atoms.

The MBIS charges are more robust than the IS charges. This is simply because MBIS pro-atoms have fewer degrees of freedom than IS pro-atoms, in line with previous observations. \cite{verstraelen2012}

The restraints in the RESP method only have a marginal impact on the robustness compared to MSK (the same method without restraints), showing that the restraints do not meet their purpose while they may cause a very poor fit to the ESP (see Fig.\ \ref{fig:esptab}). The HLY method, a rather recent ESP-fitting method with a more carefully constructed cost function, does not guarantee robust results either.

QTAIM charges are among the most robust in our test, which is in line with previous studies assessing the transferability of QTAIM results. \cite{devereux2009, popelier2015} Still, several Hirshfeld variants, such as CM5, MBIS and DDEC4 are comparably robust.

\subsection{Pareto analysis}
\label{subsec:pareto}

\begin{figure}
    \includegraphics{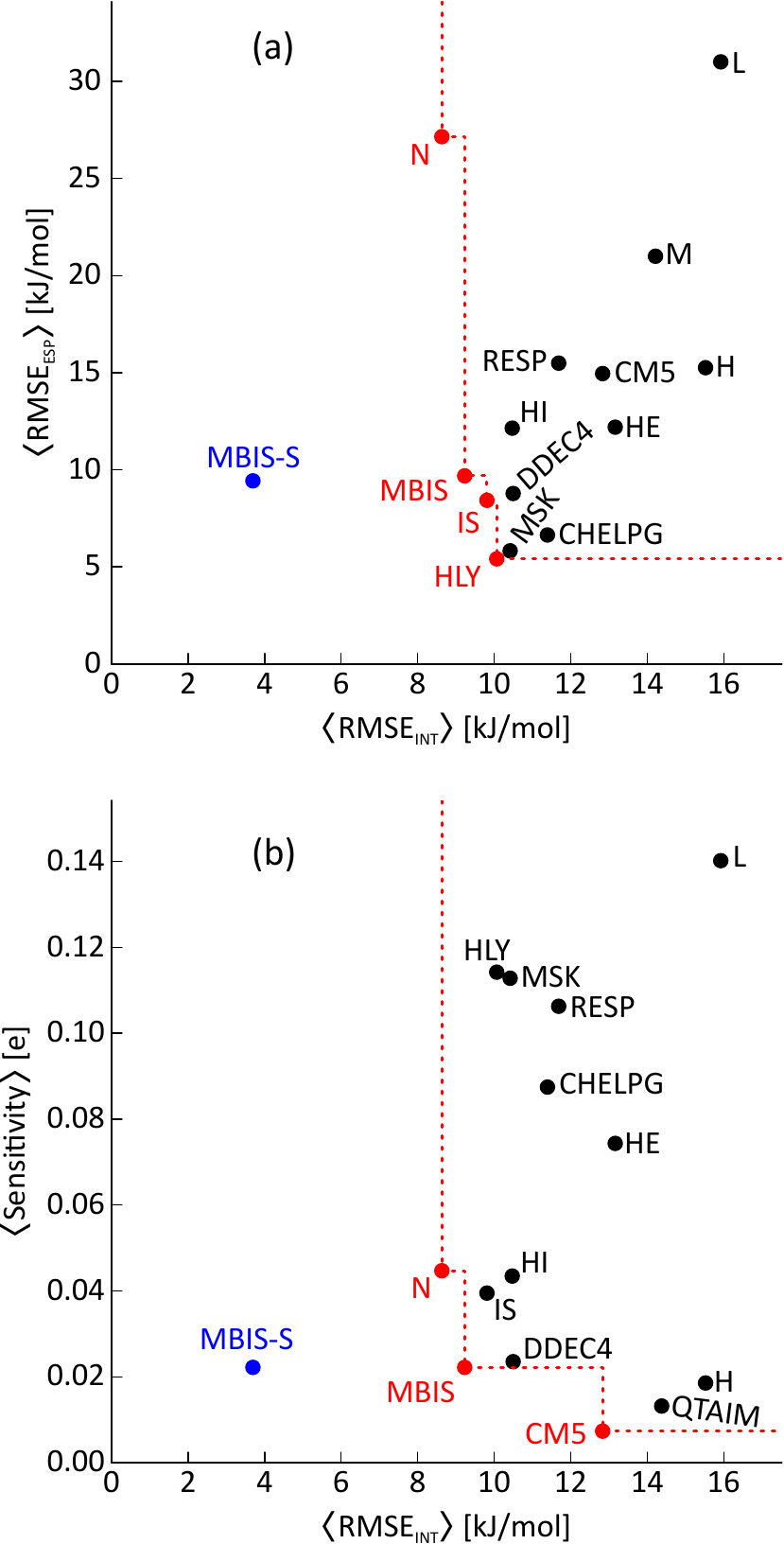}
    \caption{Pareto plots showing the trade-offs between different desirable properties of AIM methods: (a) The quality of the ESP versus the accuracy of electrostatic interactions and (b) the sensitivity of the atomic charges versus the accuracy of the electrostatic interactions. Red datapoints are on the Pareto front for all models using just point charges. The MBIS-S method goes beyond point charges by introducing a model for the distributed valence electron density.}
    \label{fig:pareto}
\end{figure}

In our comparative analysis, we have considered three main criteria that atomic charges should meet for the development of force fields: accuracy of the ESP, accuracy of electrostatic interactions and robustness. Ideally, an AIM method should combine all these features, especially the last two. The Pareto plots in Fig.\ \ref{fig:pareto} visualize the trade-offs between different criteria discussed in the previous subsections. It is unavoidable that some subjective choices slightly affect the Pareto analysis, such as the selected molecules in the datasets, their classification into groups, the ESP cost function, etc. Nevertheless, some clear trends can be observed.

Fig.\ \ref{fig:pareto}a compares the average RMSE$_\text{ESP}$ over all groups in Fig.\ \ref{fig:esptab} (Y-axis) to the average RMSE of the weak and medium electrostatic interactions from Fig.\ \ref{fig:inttab} (X-axis). The strong electrostatic interactions are not included because we found that these can never be reproduced reliably with any model using just atomic monopoles. The datapoint for QTAIM was omitted due to its excessively large average RMSE$_\text{ESP}$. The Pareto front only considers genuine point-charge models. Obviously, the MBIS-S method performs far better for electrostatic interactions as it goes beyond the simple point-charge model. This figure mainly shows that an accurate electrostatic potential does not guarantee accurate electrostatic interactions, and vice versa. If both qualities are of interest, the methods N, MBIS, IS and HLY are Pareto optimal. Obviously, HLY is Pareto optimal as its cost function was used to compute $\langle\text{RMSE}_\text{ESP}\rangle$, which is the reason for its advantage over MSK and CHELPG.

Fig.\ \ref{fig:pareto}b uses the same X-axis as figure \ref{fig:pareto}a but has the average of the three sensitivity values from Fig.\ \ref{fig:senstab} on the Y-axis. The datapoint for the Mulliken method was omitted due to its excessively large average sensitivity value. The Pareto-optimal point-charge models are N, MBIS and CM5. Again, when going beyond point charges, MBIS-S has a very attractive performance. The poor performance of ESP-fitted charges in Fig.\ \ref{fig:pareto}b is striking. The RESP method has been the method of choice in the development of many force field models, most notably in the AMBER community. Our results indicate that RESP and other ESP fitting methods are relatively poor methods for modeling electrostatic interactions in force fields.

\section{Application to density-dependent dispersion models}
\label{sec:disp}

Several dispersion models, typically used to correct DFT calculations, make use of the Hirshfeld partitioning method to estimate AIM polarizabilities. The polarizability of atom $A$ in a molecule, $\alpha_A$, is obtained by rescaling the experimental value of the free neutral atom using the third radial moment of AIM density, $\rho_A(\mathbf{r})$:
\begin{equation}
    \alpha_A = \alpha_{A,\text{free}} \frac{\langle r^3 \rangle_A}{\langle r^3 \rangle_{A,\text{free}}},
\end{equation}
where
\begin{equation}
    \langle r^3 \rangle_A = \int d\mathbf{r} |\mathbf{r} - \mathbf{R}_A|^3 \rho_A(\mathbf{r})
\end{equation}
and similarly for $\langle r^3 \rangle_{A,\text{free}}$. These rescaled polarizabilities are used in various methods \cite{becke2007, tkatchenko2009, steinmann2011} to obtain environment-specific atomic $C_6$ coefficients. The original Hirshfeld method is most often used in this context. The Iterative Hirshfeld method is sometimes used instead \cite{steinmann2010} and found to improve dispersion-corrected DFT calculations for ionic systems. \cite{bucko2013, bucko2014} Density-based dispersion models are not only used for correcting DFT calculations but were recently also employed in force-field development. \cite{cole2016} In this section, we will directly compare the accuracy of molecular $C_6$ coefficients when the Tkatchenko-Scheffler method is used in combination with different Hirshfeld variants. \cite{tkatchenko2009} We expect that similar results can be obtained with the Exchange-hole dipole model (XDM) \cite{becke2007} and related approaches such as dDsC. \cite{steinmann2011}

The expectation value $\langle r^3 \rangle_{A,\text{free}}$ can be computed in two different ways. One may use the symmetry-broken ground state density of the free atom or one may constrain the atom to be spherically symmetric and closed-shell. (In both cases, the same level of theory is used as for the molecule.) The second  choice is the most common in the context of dispersion models. For some elements however, this results in higher-energy states with fractionally occupied orbitals. For the sake of consistency, compatible choices are made when computing the reference atoms for the (Iterative) Hirshfeld method. In this section, we consider in total six variants of the TS dispersion model, using three different partitioning methods: Hirshfeld (H), Iterative Hirshfeld (HI) and MBIS. For each partitioning method, ground state reference atoms (GS) or spherical closed-shell atoms (SCS) are used. The molecular $C_6$ coefficients are also computed with Grimme's D3 model. \cite{grimme2010}

The dispersion models are tested with the database of $C_6$ coefficients for dimers of neutral molecules by Tkatchenko and Scheffler, \cite{tkatchenko2009} which are derived from experimental dipole oscillator strengths. B3LYP/6-311+G(2df,p) densities \cite{becke1993b, ditchfield1971} are computed for all molecules in this database and the atoms they contain. Gaussian09 \cite{g09} is used for the B3LYP calculations, except for the SCS atoms, for which a new SCF program was written. Dimers with the xenon atom are omitted because their all-electron density can only be computed properly with relativistic corrections. As explained in subsection \ref{subsec:impl}, the density cannot be written out by Gaussian09 when relativistic corrections are used.

\begin{figure}
    \includegraphics{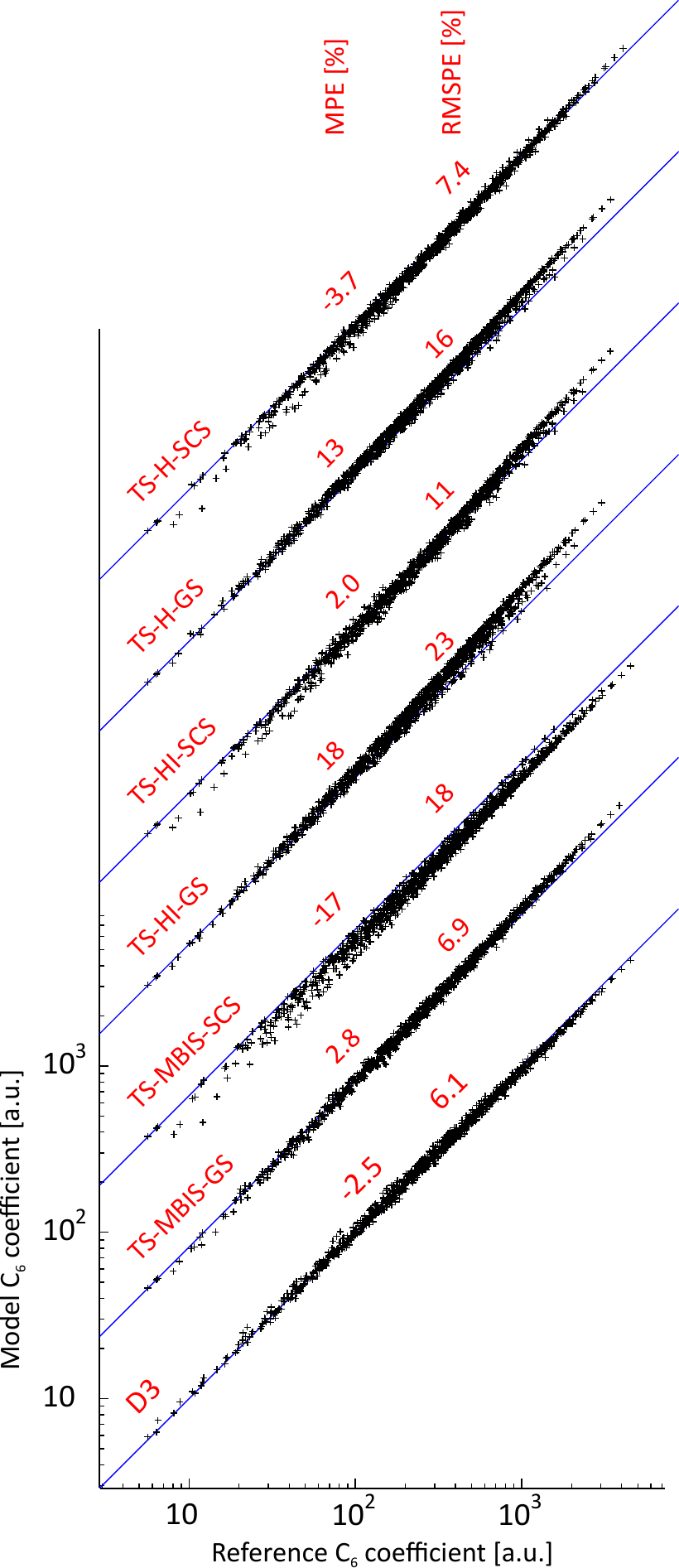}
    \caption{Scatter plots of model molecular $C_6$ coefficients versus experimental reference values. Except for the D3 results, all data points are shifted up to avoid overlap between results from different models. The mean percentage error (MPE) and the root-mean-square percentage error (RMSPE) between model and reference molecular $C_6$ coefficients are printed just above the corresponding data points.}
    \label{fig:c6}
\end{figure}

Fig.\ \ref{fig:c6} shows the scatter plots of the model $C_6$ values versus the experimental reference data, including the mean percentage errors (MPE) and the root-mean-square percentage errors (RMSPE). The original TS model corresponds to TS-H-SCS, which is one of the better variants. When using the Iterative Hirshfeld method instead (TS-HI-SCS), the RMSPE increases, indicating that the model becomes less accurate. In both H and HI variants, the use of ground-state atoms, i.e.\ TS-H-GS or TS-HI-GS, leads to a systematic overestimation of the reference $C_6$ coefficients. The situation is reversed when using MBIS partitioning, i.e.\ the dispersion model is most accurate when using ground state reference atoms (TS-MBIS-GS), while the use of spherical reference atoms (TS-MBIS-SCS) is clearly inferior. Finally, it is worth noting that Grimme's D3 model performs slightly better than any TS variant, which is impressive given that it only makes use of the nuclear coordinates and not the electron density.

We will now analyze why the TS model only works well for certain methodological combinations. First, the spherical closed-shell atoms have slightly larger (or equal) $\langle r^3 \rangle_{A,\text{free}}$ values compared to ground state atoms, as shown in Table \ref{tab:rcubed}. It turns out that, for completely different reasons, the Hirshfeld method exhibits some artifacts in the partitioning that also result in increased values of $\langle r^3 \rangle_A$, as explained below. For most molecules, these two effects balance out, except for the \ce{H2} molecule, which causes some outliers in Fig.\ \ref{fig:c6} for TS-H-SCS and TS-HI-SCS at low $C_6$ values. 

\begin{table}
    \centering
    \caption{The value of $\langle r^3 \rangle_{A,\text{free}}$ in $\mathrm{a}_0^3$ for the elements present in the molecules in the Tkatchenko-Sheffler set, \cite{tkatchenko2009} computed in two ways: using ground-state electron densities (GS) and using densities of atoms that are constrained to be spherical and closed-shell (SCS).}
    \label{tab:rcubed}
    \begin{tabular}{lddd}
    \hline
    Element &
    \multicolumn{1}{c}{$\langle r^3 \rangle_{A,\text{free,GS}}$ [$\mathrm{a}_0^3$]} &
    \multicolumn{1}{c}{$\langle r^3 \rangle_{A,\text{free,SCS}}$ [$\mathrm{a}_0^3$]} &
    \multicolumn{1}{c}{$\frac{\langle r^3 \rangle_{A,\text{free,SCS}}}{\langle r^3 \rangle_{A,\text{free,GS}}}$} \\
    \hline
    H  & 7.9   & 9.7   & 1.23  \\
    Li & 89.0  & 105.6 & 1.19  \\
    C  & 35.7  & 41.4  & 1.16  \\
    N  & 27.0  & 30.9  & 1.14  \\
    O  & 22.7  & 23.8  & 1.05  \\
    F  & 18.6  & 18.9  & 1.01  \\
    Ne & 15.4  & 15.4  & 1.00  \\
    Si & 103.0 & 114.3 & 1.11  \\
    S  & 77.0  & 79.7  & 1.04  \\
    Cl & 66.7  & 67.4  & 1.01  \\
    Ar & 57.2  & 57.2  & 1.00  \\
    Br & 97.0  & 97.8  & 1.01  \\
    Kr & 89.1  & 89.1  & 1.00  \\
    \hline
    \end{tabular}
\end{table}

\begin{figure}
    \includegraphics{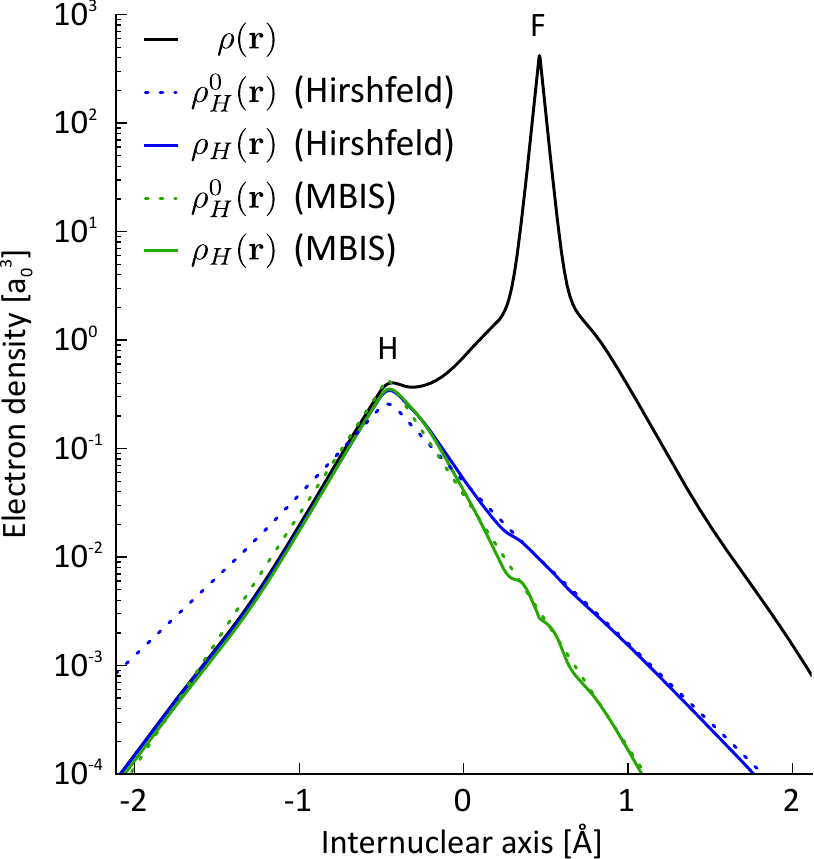}
    \caption{Total electron density (black) of hydrogen fluoride along the internuclear axis, including the AIM density (solid) and the pro-atomic density (dashed) of hydrogen for the Hirshfeld (blue) and MBIS method (green).}
    \label{fig:hf}
\end{figure}

The artifact of the Hirshfeld method is very clear in Fig.\ \ref{fig:hf}: it shows the AIM density, $\rho_A$, of the hydrogen atom in hydrogen fluoride, computed with the Hirshfeld and MBIS methods. The Hirshfeld AIM density (solid blue line) is asymmetric, with more electron density toward the fluoride. This can be understood as follows. Any variant of the Hirshfeld method exhibits the same similarity principle, \cite{ayers2000} due to Eq.\ \eqref{eq:stockholder}: the Hirshfeld AIM density will be as close as possible to that of the pro-atom. Because the density tail of isolated hydrogen (dashed blue line) is so different from that of hydrogen in HF, the Hirshfeld AIM density is aspherical with too much density in the bonding region. This accumulation of density, relatively far away from the hydrogen nucleus, leads to larger values of high radial moments, such as $\langle r^3 \rangle_A$. This is a general feature of the Hirshfeld method, also seen in other molecules. In the MBIS method, no such asymmetries are found because the parameters $\sigma_{A,i}$ in Eq.\ \eqref{eq:proslater}, i.e.\ the widths of the Slater functions, are also optimized to match the molecular electron density. By consequence, the MBIS AIM density (solid green curve) almost coincides with the corresponding pro-atom (dashed green curve) and is therefore close to symmetric.

Fig.\ \ref{fig:hf} also shows a side effect of the MBIS method: where the MBIS AIM density (solid green curve) passes through the nucleus of fluoride, some ripples can be seen, because some details in the fluoride core electron density cannot be reproduced by the Slater functions. These ripples are very local and therefore have a negligible effect on $\langle r^3 \rangle_A$.

\begin{figure}
    \includegraphics{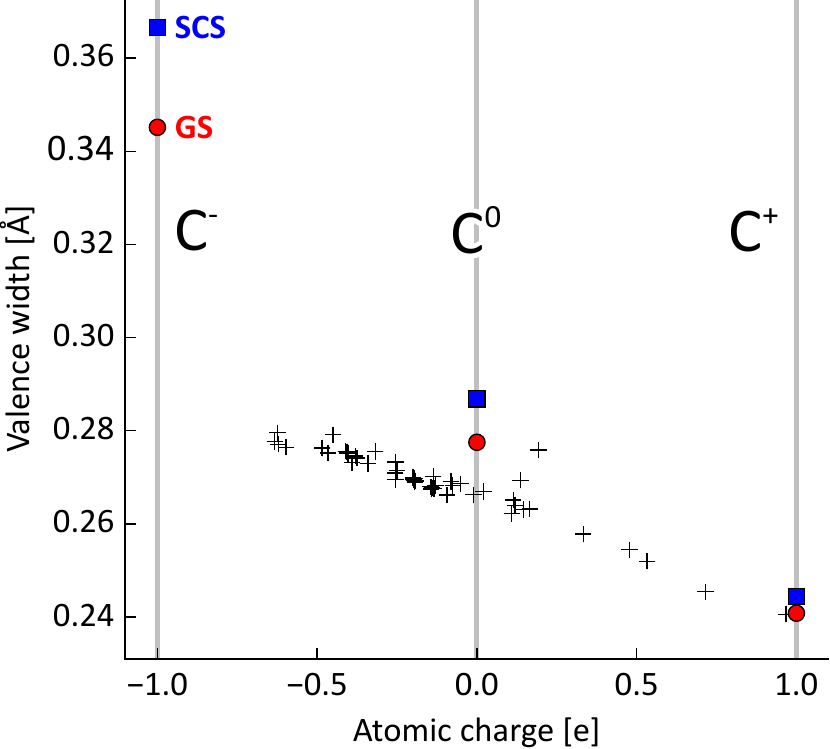}
    \caption{The MBIS \textit{valence width} ($\sigma_{A,v}$, see subsection \ref{subsec:ff}) of carbon atoms in the Tkatchenko-Sheffler set of molecules \cite{tkatchenko2009} versus their MBIS charge (black plus). The valence widths of isolated carbon atoms and ions are also included: ground states atoms (red circle) and spherical closed-shell atoms (blue square).}
    \label{fig:vw}
\end{figure}

Fig.\ \ref{fig:vw} further illustrates the mismatch between the density tails of the (Iterative) Hirshfeld pro-atoms and the molecular densities in the TS set. It shows the MBIS \textit{valence width} of the carbon atoms of all molecules in the TS set \cite{tkatchenko2009} versus their MBIS charge. This figure also includes the data points for free carbon atoms and ions, which are used as pro-atoms in the (iterative) Hirshfeld method. The trends in this figure are general: neutral atoms and anions, have density tails that decay slower (i.e.\ higher valence width) than molecular electron densities. This leads to the asymmetric (Iterative) Hirshfeld AIM densities as the one observed in Fig.\ \ref{fig:hf}. Especially when computing higher radial or multipole moments with the (Iterative) Hirshfeld method, this may result in undesirable artifacts. In previous work, the erroneous density tails of unstable anions were corrected, e.g. by computing these atoms in a Watson sphere \cite{watson1958}. For example, such corrections were used by Bu\v{c}ko \textit{et al.} \cite{bucko2013, bucko2014} in their tests of the TS-HI variant. The results in Fig.\ \ref{fig:vw} suggest that it is also advantageous to reduce the density tails of stable anions and neutral atoms.

Finally, note that a few other Hirshfeld variants were proposed, in which the density tails of the pro-atoms are optimized to match the molecular electron density, most notably the Iterative Stockholder (IS) method \cite{lillestolen2008, lillestolen2009} and some of its variants including the Gaussian ISA \cite{verstraelen2012} and BS-ISA+DF. \cite{misquitta2014} 

\section{Conclusions and outlook}
\label{sec:concl}

The MBIS method is a new density-based AIM method that is particularly suitable for the development of efficient and relatively accurate electrostatic force-field models. MBIS belongs to the family of Hirshfeld methods. Its pro-density is expanded in a minimal set of atom-centered s-type Slater density functions, whose parameters are fitted to a given molecular electron density by minimizing the Kullback-Leibler divergence. In that sense, it can also be interpreted as an information theory density-fitting method, where the Slater functions as such are used in applications, rather than the atoms-in-molecules densities.

The MBIS method is extensively tested for the development of electrostatic force-field models. When it is just used for the purpose of deriving atomic charges, it is one of the best methods available to date, in terms of robustness and accuracy of the electrostatic interactions. When the MBIS Slater functions are used to describe the valence electron density in a force field, the error on the electrostatic interactions can be reduced by 50\%, if the electrostatic interaction is not too strong. This is a computationally efficient approach to describe the so-called penetration effect, i.e.\ the deviation of interatomic electrostatic interactions from that of point charges, when the atomic densities begin to overlap. MBIS is also useful beyond the scope of frozen-density electrostatics,  e.g.\ when modeling dispersion interactions, or to analyze density fluctuations that a polarizable force field should reproduce.

In future work, we will focus on improving our method, its implementation and more applications in different areas. The obvious methodological improvement is a better model for the pro-molecular density, e.g. by including atomic multipoles \cite{misquitta2014, ohrn2016} or by making it compatible with pseudo-densities. An improved pro-molecule model should not merely result in a better fit to a given electron density; one should also avoid too many degrees of freedom for the sake of robustness. Moreover, the use of an improved pro-density model in force fields should remain computationally efficient.

\begin{acknowledgement}
The authors would like to thank F. De Proft and P. Bultinck for inspiring discussions and useful suggestions. T.V., S.V.\ and L.V.\ acknowledge the Foundation of Scientific Research - Flanders (FWO). T.V., L.V., M.W.\ and V.V.S.\ thank the Research Board of Ghent University (BOF), and BELSPO in the frame of IAP/7/05 for their financial support. V.V.S.\ acknowledges the European Research Council for funding the European Union’s Horizon 2020 research and innovation programme [consolidator ERC grant agreement no. 647755 -- DYNPOR (2015--2020)]. The computational resources and services used were provided by Ghent University (Stevin Supercomputer Infrastructure). F.H.-Z.\ was supported by a Vanier-CGS fellowship and a Michael Smith Foreign Study Supplement from the National Sciences and Engineering Research Council of Canada (NSERC); P.W.A.\ was supported by a Discovery Grant and a E.W.R.\ Steacie Fellowship form NSERC.
\end{acknowledgement}

\begin{suppinfo}
    Atom types in MIL-53(Al), chabazite and ozone.
    Description of the ZG237 set of molecular dimers and Cartesian coordinates of the MP2/cc-pVTZ+CP optimized dimer geometries in this set.
\end{suppinfo}

\bibliography{references}

\end{document}